\documentclass{article}
\usepackage{typearea}
\typearea{12}
\usepackage{amsmath, amssymb}
\usepackage{amsfonts}
\usepackage{mathrsfs}
\usepackage{cite}
\usepackage[dvipdfmx]{graphicx, xcolor}
\usepackage{array, booktabs}
\usepackage{comment}
\usepackage{caption}


\begin{document}

\title{\bf \large Leptogenesis from composite singlets}
\author{\normalsize Nobuki Yoshimatsu}
\date{\it \footnotesize Supreme School for Advanced Education\\1-5-4\ Nishi Takamatsu, Wakayama-shi, Wakayama 641-0051, Japan\\ 
Chiben Gakuen Wakayama Junior/Senior High School \\ 2066-1\ Fuyuno, Wakayama-shi, Wakayama 640-0392, Japan\\
{\footnotesize E-mail address:\ nyoshimatsu260@gmail.com}}

\maketitle
\begin{center}
Abstract
\end{center}
{\small We argue that dual singlets accommodate the thermal leptogenesis in the context of metastable supersymmetry breaking. 
This framework suggests that the reheating temperature is as low as $\mathcal{O}(10^4)$ GeV, and besides the neutrino masses, solely induced from the dual fermions, are present at $\mathcal{O}(0.1)$ eV. We also expect that the unstable gravitino, though long-lived, may cause a clear signal of the neutrino flux at $\mathcal{O}(10)$ MeV.}
{\flushleft{\ }}
\section{Introduction}

The baryon asymmetry of the universe is an interesting topic in particle physics. According to the CMS report \cite{WMAP:2010qai}, the baryon abundance (denote $\eta_{B}$) to that of photons (denoted $\eta_{\gamma}$) is estimated as 
\begin{equation}
\dfrac{\eta_{B}}{\eta_{\gamma}} \simeq 6 \times10^{-10}. \label{eq.1}
\end{equation}
Regarding its origin,  thermal leptogenesis provides an elegant mechanism in which the lepton number from the right-handed neutrino decay is fractionally converted into the baryon number via a sphaleron process \cite{Fukugita Yanagida, Buchmuller:2002rq, FY2, FY3}.
The hypothesis, however, entails a high reheating temperature of $T_{RH}\gtrsim 10^9$ GeV caused by the heavy right-handed neutrino(s), which encounters the gravitino overproduction under the circumstance of $m_{3/2} \lesssim 1$ GeV.

\noindent
In this letter, we demonstrate that the dual singlet bosons, present in the Intriligator-Seiberg-Shih (ISS) type of dynamical SUSY breaking (DSB) \cite{dual}, accommodate the leptogenesis through a pseudo Nambu-Goldstone boson (NGB) loop and the discrete $R$ symmetry violating interaction with the minimal supersymmetric standard model (MSSM) sector.
At this point, the authors of \cite{Grossman:2003jv, DAmbrosio:2003nfv, Grossman:2004dz} proposed the so-called Soft Leptogenesis, where the two real sneutrinos, that belong to the supermultiplet of a single generation, have the degenerate mass originating from the grand unification theory (GUT) scale, though the soft mass terms induce the mass splitting, thus leading to the thermal leptogenesis with $T_{RH}=10^6-10^8$ GeV. In contrast, we show that even without invoking the GUT (and irrespective of the flavor physics), the SUSY breaking necessarily makes a pair of dual singlet bosons (both of which are the complex scalars) mass degenerate, whereas its degeneracy can be lifted via higher-order couplings. We subsequently illustrate that $T_{RH}$ is rendered $\mathcal{O}(10^4)$ GeV, and  the dual singlet fermions induce the neutrino masses at $\mathcal{O}(0.1)$ eV in the absence of the GUT as well\cite{Biondini:2017fut}. Further, our model suggests that the neutrino flux, emitted from the long-lived gravitino, may exhibit a clear excess at $\mathcal{O}(10)$ MeV, while the extragalactic gamma-ray flux is left observationally acceptable.\\

\section{Explicit model based on $SU(3)$ gauge group}

\noindent
Let us illustrate an explicit model based on the ISS mechanism as displayed in Table $1$. Here, 
we assume the hidden $SU(3)$ gauge group of $N_F=4$ with a super Yang-Mills sector (the gauge field strength superfield is denoted $W^{\alpha}$) and a $Z_{4R}$ symmetry: 
\begin{center}
\begin{tabular}{c|c|c}
\multicolumn{3}{c}{Table 1. Matter content/Charge assignment}
\\ \toprule \addlinespace[2pt]
&    $Q^1, Q^2, Q^3, Q^4$ & $\bar{Q}^1, \bar{Q}^2, \bar{Q}^3, \bar{Q}^4$ \\ \hline
\addlinespace[2pt] $SU(3)$ & $\bf{3}$ & $\bf{\bar{3}}$  \\ \hline
\addlinespace[2pt] $Z_{4 R}$ & $2,\ 1,$  $1,\ 1$ & $2,\ -1,$  $-1,\ -1$ \\ \bottomrule
\end{tabular}
\end{center}
Among the most general superpotential consistent with all the symmetries are the following couplings \cite{planck-suppressed-dual-mass}:
\begin{align}
W_{el}  \supset &\  \lambda_{11} m \bar{Q}^1Q^1  + \sum_{i,j=2}^4 \lambda_{ij} m \bar{Q}^iQ^j \notag \\
&+\sum_{i,j=2}^4 \eta_{ij}\left(\dfrac{m}{M_{pl}}\right)  \dfrac{\bar{Q}^1Q^1 \bar{Q}^iQ^j}{M_{pl}} \notag \\& +\sum_{i,j,k,l=2}^4 \kappa_{ijkl}\left(\dfrac{m}{M_{pl}}\right) \dfrac{\bar{Q}^iQ^j \bar{Q}^kQ^l}{M_{pl}} \label{deltaQ}
\end{align} 
where $m$ is defined as
\begin{equation}
m \equiv \dfrac{\left<Tr W^{\alpha}W_{\dot{\alpha}}\right>}{M_{pl}^2}.
\end{equation}
 Further, we introduce additional interactions with the SM leptons and the Higgs superfields that violate the $Z_{4R}$ symmetry:
\begin{gather}
W^{RV} = \sum_{n=1}^3\  \sum_{i, j=2}^4 y_{ij}^n\dfrac{\bar{Q}^i Q^j H_u L_n}{M_{pl}}, \label{sm-hidden}
\end{gather}
where $n$ denotes the SM generation index. After diagonalizing the $\bar{Q}^{i} Q^j$ mass matrix (the eigenvalue is denoted $m_i$ which is ordered $m_1>m_2>m_3>m_4$) and below the dynamical scale (denoted $\Lambda$), the superpotential in Eqs.(\ref{deltaQ}) and (\ref{sm-hidden}) is written in terms of the dual theory:
\begin{align}
& W_{dual} =\sum_{i,j=1}^4 \bar{b}_iS^{ij}b_j-\dfrac{\det{S^{ij}}}{\Lambda} +  \sum_{i=1}^4 m_i \Lambda S^{ii}  \\ &+\sum_{i,j=2}^4   \eta_{ij}\dfrac{m \Lambda^2}{M_{pl}^2} S^{11} S^{ij} +\sum_{i,j,k,l=2}^4 \kappa_{ijkl}  \dfrac{m \Lambda^2}{M_{pl}^2} S^{ij} S^{kl} \label{tiny-mass-difference} \\ &+W^{RV}_{dual} \label{composite-fermion-rev}
\end{align}
with
\begin{align}
W^{RV}_{dual}= & \sum_{n=1}^3 \ \sum_{i, j=2}^4 y_{ij}^n\dfrac{\Lambda S^{ij} H_u L_n}{M_{pl}}. \label{dual-yukawa}
\end{align}
Here, $S^{ij}, \bar{b}_i, b_i$ are defined as
\begin{equation}
S^{ij}=\dfrac{\bar{Q}^i Q^j}{\Lambda}, \   \bar{b}_i=\epsilon_{ijkl} \dfrac{\bar{Q}^j \bar{Q}^k \bar{Q}^l}{\Lambda^2}, \   b_i= \epsilon_{ijkl} \dfrac{Q^j Q^k Q^l}{\Lambda^2}
\end{equation}
with $\epsilon_{ijkl}$ standing for the totally anti-symmetric tensor. 
At metastable SUSY breaking vacuum, it follows that 
\begin{gather}
\left<b_1\right>=\left<\bar{b}_1\right>=\sqrt{m_1 \Lambda}, \label{ngb}\\
 F_{S^{ii}}=m_i \Lambda,\ (i=2-4),\ m_{3/2}= \dfrac{F_{total}}{\sqrt{3} M_{pl}},
\end{gather}
where
\begin{equation}
F_{total}^2=\sum_{i=2}^4 F_{S^{ii}}^2.
\end{equation}
Besides, the $S^{ij} \ (i, j=2-4)$ bosons acquire their mass via $\bar{b}_i, b_i$ ans $S^{ij}$ loops:
\begin{align}
V_{1 loop}  = & \sum_{i, j=2}^4 \left(M^{loop}_{ij}\right)^2 \left(\left|S^{ij} \right|^2 + \left|S^{ji} \right|^2\right)\ \  (i\neq j), \notag \\ 
&+\sum_{i=2}^4 \left(M^{loop}_{i}\right)^2 \left|S^{ii} \right|^2, \label{degenerate-loop-induced}
\end{align}
where
\begin{gather}
\left(M^{loop}_{ij}\right)^2 \equiv  \dfrac{\left(m_{i}^{2} + m_{j}^{2} \right)\Lambda^2}{48 \pi^2 m_1\Lambda}, \notag \\
\left(M^{loop}_{i}\right)^2  \equiv \dfrac{2 m_{i}^{2}\Lambda^2}{48 \pi^2 m_1\Lambda}.
\end{gather}
(See Appendix A for detailed calculations.) In what follows, we demonstrate that the degeneracy is lifted between $S^{ij}$ and $S^{ji} \ (i \neq j)$ due to Eq.(\ref{tiny-mass-difference}), which generates the lepton number.  \\

\section{Generation of lepton number}

\noindent
Let us evaluate the lepton asymmetry owing to the $S^{ij} \ (i, j=2-4)$ boson decay. Eq.(\ref{dual-yukawa}) leads to a decay mode into $\tilde{h}_u,  l_n$: 
\begin{align}
& \mathcal{L}  \supset \sum_{n=1}^3 \sum_{i,j=2}^4 \dfrac{y_{ij}^n \Lambda}{M_{pl}} S^{ij} \tilde{h}_u l_n +h.c..
\end{align} 
Meanwhile, another mode appears via Eq.(\ref{ngb}). As noted by \cite{M.N}, there is an NGB:
\begin{equation}
Im\left(\dfrac{b_1-\bar{b}_1}{\sqrt{2}}\right) \equiv \phi
\end{equation}
 from the spontaneous breaking of $U(1)_B$ (although $\phi$ possibly acquires its mass through the higher-order coupling as discussed later).
The $S^{ij}$ bosons are found to interact with $\phi$:
\begin{gather}
V \supset \left|\dfrac{\partial W_{dual}}{\partial S^{11}}\right|^2 \supset \sum_{i,j=2}^4 \eta_{ij} \dfrac{\Lambda^2}{M_{pl}} S^{ij} \bar{b}_1 b_1+h.c., \notag \\ \supset \sum_{i,j=2}^4 \dfrac{\eta_{ij} \Lambda^2}{2! M_{pl}} S^{ij} \phi^2+h.c.. \label{inverse-decay}
\end{gather} 
At this point, it should be noted that the $Z_{4R}$ symmetry forbids the $\bar{Q}^iQ^j \bar{Q}^kQ^l/M_{pl}$ coupling in Eq.(\ref{deltaQ}). Otherwise, the dual singlet bosons would preferentially decay into two $\phi$s, resulting in much suppression of the generation of the lepton number \cite{S-two-phi}. We conduct further analyses, based on the mass eigenstates of $S^{ij}$ bosons, which are referred to as $\chi^a_{\pm}\ (a=1-3)$ and $\chi^{ii}\ (i=2-4)$.  Also, $\kappa_{ijkl}$ is set for simplicity:  
\begin{gather}
\kappa_{ijkl} = \begin{cases} \kappa_1, \ \ \ (i,j)=(k,l) \notag \\ \kappa_2, \  \ \ (i,j) \neq (k,l). \end{cases}
\end{gather}
 One then obtains each eigenvalue: 
\begin{gather}
 \begin{pmatrix} m^2_{\chi^{a}_{+}} \\ m^2_{\chi^{a}_{-}} \end{pmatrix}=\begin{pmatrix} \left(M^{loop}_{ij}\right)^2  \\  \left(M^{loop}_{ij}\right)^2\end{pmatrix} +\begin{pmatrix} \Delta m^2_{\chi^{a}_+} \\ \Delta m^2_{\chi^{a}_-} \end{pmatrix}, \\
m^2_{\chi^{ii}} \simeq \left(M^{loop}_{i}\right)^2,
\end{gather}
where
\begin{gather}
\Delta m^2_{\chi^{a}_{+}} \equiv \left[U^T\Delta \mathcal{M}^2_{\chi^a} U\right]_{11},
\  \Delta m^2_{\chi^{a}_{-}} \equiv \left[U^T\Delta \mathcal{M}^2_{\chi^a} U\right]_{22}, \label{u1}
\end{gather}
and 
\begin{align}
&\Delta \mathcal{M}^2_{\chi^a}= \left(\dfrac{m \Lambda^2}{M_{pl}^2}\right)^2 \notag \\ & \begin{pmatrix} \left|\eta_{ij} \right|^2 + \left|\kappa_1 \right|^2+8 \left|\kappa_2 \right|^2 & \eta_{ij} \eta^{\ast}_{ji}+ \kappa^{\ast}_1 \kappa_2+ \kappa_1 \kappa^{\ast}_2 +7\left|\kappa_2 \right|^2\\ \eta_{ij}^{\ast} \eta_{ji} + \kappa_1 \kappa^{\ast}_2+ \kappa_1^{\ast} \kappa_2+7\left|\kappa_2 \right|^2& \left|\eta_{ji} \right|^2+ \left|\kappa_1 \right|^2+8 \left|\kappa_2 \right|^2 \end{pmatrix}. \label{matrix-1}
\end{align}
Here, we take into consideration $\left(m^{loop}_{a}\right)^2, \left(m^{loop}_{ii}\right)^2 \gg \Delta m^2_{\chi^{a}}$.
Accordingly, the coupling constants $\eta_{{ij}}, y_{ij}^n$ are transformed into $\xi_{a_{\pm}}, y^{\prime n}_{a_{\pm}}$: 
\begin{equation}
\begin{pmatrix} \xi_{a_+} \\ \xi_{a_-} \end{pmatrix}= U^T \begin{pmatrix} \eta_{ij} \\ \eta_{ji} \end{pmatrix},\ \ \begin{pmatrix} y^{\prime n}_{a_+} \\y^{\prime n}_{a_-}\end{pmatrix} =U^T \begin{pmatrix} y_{ij}^n \\ y_{ji}^n \end{pmatrix}, 
\end{equation}
where $U$ is the $2 \times2 $ unitary matrix that diagonalizes $\Delta \mathcal{M}^2_{\chi^a}$, and each subscript is combined as in $(a, ij)=(1, 23), (2, 24), (3, 34)$. %At this point, as in the Soft Leptogenesis  \cite{Grossman:2003jv, DAmbrosio:2003nfv, Grossman:2004dz}, 
 In Tables 2.1 and 2.2, we enumerate what each parameter stands for. 
\begin{center}
\begin{tabular}{c|c|c}
\multicolumn{3}{c}{Table 2.1 Notation of dual bosons and its mass}
\\ \toprule \addlinespace[2pt]
current state & eigenstate & eigen mass \\ \hline
\addlinespace[2pt]  $S^{ij}$ &  $\chi^{a}_{\pm}$&$m_{\chi^{a}_{\pm}}$ \\ \bottomrule
\end{tabular}
\end{center}
\begin{center}
\begin{tabular}{c|c}
\multicolumn{2}{c}{}
\\ \toprule \addlinespace[2pt]
loop-induced mass of $S^{ij}$ &  extra contribution to squared-mass of $\chi^{a}_{\pm}$ \\ \hline
\addlinespace[2pt] $M^{loop}_{ij}$  &  $\Delta m^2_{\chi^{a}_{\pm}}$    \\ \bottomrule
\end{tabular}
\end{center}
\begin{center}
\begin{tabular}{c|c}
\multicolumn{2}{c}{Table 2.2 Notation of coupling constant}
\\ \toprule \addlinespace[2pt]
current state &  eigenstate \\ \hline
$y_{ij}^n$,\ \ $\eta_{ij}^n$ &  $y^{\prime n}_{a_{\pm}}$,\ \ $\xi_{a_{\pm}}$ 
 \\ \bottomrule
\end{tabular}
\end{center}
Eventually, the Lagrangian density of relevance is reduced to the form of
\begin{equation} \mathcal{L} \supset \sum_{i,j=2}^4 \dfrac{\xi_{a_{\pm}} \Lambda^2}{2! M_{pl}} \chi^{a}_{\pm} \phi^2+ \sum_{n=1}^3 \sum_{i,j=2}^4 \dfrac{y^{\prime n}_{a_{\pm}} \Lambda}{M_{pl}} \chi^{a}_{\pm}  \tilde{h}_u l_n +h.c.. 
\end{equation}
We then address the physical phases of the dimensionless coupling constants concerned. Note that $\eta_{ij}$ becomes real positive-valued after redefining $S^{ij}$. Besides, postulated that $y^n_{34}=y^n_{43}$ for simplicity (we go on our analysis under this circumstance), both $y^n_{34}$ and $y^n_{43}$ can be made the same sort of parameters, via the redefinition of $L_n$ (apart from the other $y^n_{ij}$ that have less implications on the generation of the lepton number).
To summarize, the CP violation solely emerges through the parameter:
\begin{equation}
\phi_{cp} \equiv Arg\left(\eta_{34} \eta^{\ast}_{43}+ \kappa^{\ast}_1 \kappa_2+ \kappa_1 \kappa^{\ast}_2 +7\left|\kappa_2 \right|^2\right)
\end{equation}
 in our framework.
We are in a position to derive the formula for the lepton asymmetry. The decay rates of relevance are given as follows:
\begin{gather}
\Gamma_{\chi^{a}_{\pm} \rightarrow \tilde{h}_u,  l_n} \simeq \dfrac{\left|y^{\prime}_{a^n_{\pm}} \right|^2}{8 \pi} \left(\dfrac{\Lambda}{M_{pl}}\right)^2  m_{\chi^{a}_{\pm}}, \\
\Gamma_{\chi^{a}_{\pm} \rightarrow \phi \phi} \simeq \dfrac{\left|\xi_{a_{\pm}} \right|^2}{8 \pi} \left(\dfrac{m \Lambda^2}{M_{pl}^2}\right)^2 \dfrac{1}{m_{\chi^a_{\pm}}}.
\end{gather}
We thus find that $\chi^{a}_{\pm}$ decays yield the lepton number through the wave function renormalization from a $\phi$ loop and a $\chi^a_{\mp}$ propagator \cite{Covi:1996wh}. 
Allowing for the quasi-degeneracy between $m_{\chi^{a}_{+}}$ and $m_{\chi^{a}_{-}}$, we deduce the $CP$ violating parameter:
\begin{align}
&\varepsilon_{l_n}^{\chi^{a}_{\pm}}  \simeq  \dfrac{-2 \cdot 2}{16 \pi} \cdot \dfrac{1}{2} \cdot BR \left(\chi^{a}_{\pm} \rightarrow \tilde{h}_u,  l_n \right)\  Im\left(\mathcal{I}_{a_{\pm}}\right) \notag \\ & \times \dfrac{\left(\Delta m^2_{\chi^{a}_{\pm}}-\Delta m^2_{\chi^{a}_{\mp}}\right)\left(\dfrac{m \Lambda^2}{M_{pl}^2}\right)^2}{\left(\Delta m^2_{\chi^{a}_+}-\Delta m^2_{\chi^{a}_-}\right)^2+ m^2_{\chi^a_{\mp}} \left(\Gamma^{total}_{\chi^a_{\mp}}\right)^2}
\end{align}
where $BR \left(\chi^{a}_{\pm} \rightarrow \tilde{h}_u,  l_n \right)$ is the branching ratio involved: 
\begin{equation}
 BR \left(\chi^{a}_{\pm} \rightarrow \tilde{h}_u,  l_n \right) = \dfrac{\sum_{n=1}^3\Gamma_{\chi^{a}_{\pm} \rightarrow \tilde{h}_u,  l_n}}{\Gamma^{total}_{\chi^a_{\pm}}}
\end{equation}
and $\Gamma^{total}_{\chi^a_{\pm}}, \ \mathcal{I}_{a_{\pm}}$ are written as
\begin{gather}
\Gamma^{total}_{\chi^a_{\pm}}=\sum_{n=1}^3\Gamma_{\chi^{a}_{\pm} \rightarrow \tilde{h}_u,  l_n}+\Gamma_{\chi^{a}_{\pm} \rightarrow \phi \phi}, \\
\mathcal{I}_{a_+}=\dfrac{\sum_{n=1}^3  y_{a_{+}}^{\prime n} y_{a_-}^{\prime n \ast} \xi_{a_+} \xi_{a_-}^{\ast}}{\sum_{n=1}^3 \left|y_{a_{+}}^{\prime n}\right|^2},\\ 
\mathcal{I}_{a_-}=\dfrac{\sum_{n=1}^3  y_{a_{+}}^{\prime n \ast} y_{a_-}^{\prime n} \xi_{a_+}^{\ast} \xi_{a_-}}{\sum_{n=1}^3 \left|y_{a_{-}}^{\prime n}\right|^2}.
\end{gather}
We also note that $Arg\left(\mathcal{I}_{a_+}\right) =2\phi_{cp}$ and  $Arg\left(\mathcal{I}_{a_-}\right)= -2\phi_{cp}$. \\

\section{Baryon asymmetry}

\noindent
We demonstrate that the baryon asymmetry considerably emerges from the sphaleron process. 
 Therein, we assume a moderate hierarchy among $m_i\ (i=1-4)$ to go on further estimation. 
Under the circumstance of $K_{\chi^{a}_{\pm}}^{\tilde{h}_u, l_k} \ll 1$, i.e. the weak washout regime, the final efficient factor is reduced to the following form:
\begin{align}
 \kappa_f \left(\chi^a_{\pm} \right) \simeq   \dfrac{2}{15}  K_{\chi^a_{\pm}}^{\tilde{h}_u, l_k} \times  \left(\dfrac{T_{RH}}{m_{\chi^{a}_{\pm}}}\right)^5,
\end{align}
for $T_{RH} <\ m_{\chi^{a}_{\pm}}$\cite{Giudice:2003jh, note1}.  Here, $K_{\chi^a_{\pm}}^{\tilde{h}_u, l_k}$ 
is conventionally defined as
\begin{align}
 K_{\chi^{a}_{\pm}}^{\tilde{h}_u, l_k} \equiv \dfrac{\Gamma_{\chi^{a}_{\pm}}^{\tilde{h}_u, l_k}}{H\left(m_{\chi^{a}_{\pm}}\right)}, 
\end{align}
and the Hubble parameter is given by
\begin{gather}
H\left(T \right) \simeq  62.8\times \left(\dfrac{g_{\ast}\left(T \right)}{228.75}\right)^{1/2} \cdot \dfrac{T^4}{T^2_{RH} M_{pl}},
\end{gather}
where 
$g_{\ast}(T)$ is the degree of freedom of radiative particles, and we used $g_{\ast}\left(m_{\chi^{a}_{\pm}}\right) \simeq g_{\ast}\left(T_{RH}\right) =228.75$. Choosing $m=m_1=10^{10.7}$ GeV, $m_2=10^{6}$ GeV, $m_3=10^{5.3}$ GeV, $m_4=10^{5}$ GeV, $\Lambda=10^{11.7}$ GeV, and setting $|\kappa_1|=1.0,\ |\kappa_2|=0.3$, $\eta_{ij}=1.0,\ \eta_{ji}=0.5\ (i <j),\ \left|y^n_{ij}\right|= \left|y^n_{ji}\right|\ (i \neq j)$,
we obtain the total baryon asymmetry:
\begin{align}
&\dfrac{n_B}{s} \simeq \dfrac{90 \zeta\left(3\right)}{4 \pi^4 g_{\ast}} \dfrac{24+4n_H}{66+13n_H} \cdot \left(-\sum_{a=1}^3 \epsilon^{\chi^{a}_{\pm}}_{l_n} \cdot \kappa_f \left(\chi^a_{\pm}\right)\right) \notag \\ &\ \ \ \   \simeq  8.808 \times 10^{-11} \times \left(\dfrac{\sum_{n=2}^4 \left|y^n_{34} \right|^2}{0.6}\right) \cdot \left(\dfrac{T_{RH}}{3.0 \times10^{4}\ \text{GeV}}\right)^7 \left(\dfrac{\sin{2\phi_{cp}}}{1.0}\right),
\end{align}
%\begin{align}
%&\dfrac{n_B}{s} \simeq \dfrac{90 \zeta\left(3\right)}{4 \pi^4 g_{\ast}} \dfrac{24+4n_H}{66+13n_H} \cdot \left(-\sum_{a=1}^3 \epsilon^{\chi^{a}_{\pm}}_{l_n} \cdot \kappa_f \left(\chi^a_{\pm}\right)\right) \notag \\ & \simeq 1.48 \times 10^{-9} %\sum_{n=2}^4 
%\left|y^n_{34} \right|^2  \times \left(\dfrac{T_{RH}}{3.0 \times10^{4}\ \text{GeV}}\right)^7 \sin{2\phi_{cp}}, 
%\end{align}}
where $n_H$ is the number of the Higgs doublets that equals $2$ in the MSSM, and we used the unitary matrix $U$ of Eq.(\ref{u1}), expressed as
\begin{equation}
U= \begin{pmatrix} 0.804 & 0.595 \\ -0.595 e^{-i \phi_{cp}} & 0.804 e^{-i \phi_{cp}}\end{pmatrix}.
\end{equation} 
%with
%\begin{equation}
%\xi_{a_+}\xi_{a_-}^{\ast}=-0.141 e^{i \phi_{cp}},\ \ y^{\prime n}_{a_+}y^{\prime n \ast}_{a_-}=0.405 e^{i \phi_{cp}}.
%\end{equation}
At this point, we stress that $\chi^3_{\pm}$ are nearly mass-degenerate:
\begin{equation}
\dfrac{m_{\chi^3_+}- m_{\chi^3_-}}{m_{\chi^3_+}} =\mathcal{O}(10^{-15}),
\end{equation}
so that both almost simultaneously begin to decay, and then we observe that the $\chi^3_{\pm}$ decays are relevant to the generation of the lepton asymmetry. (See Appendix B for the numerical evaluation based on the Boltzmann equations concerned.)
%, and also emphasize that each inverse decay rate, the magnitude of which is $K_{\chi^{a}_{\pm}}^{\tilde{h}_u, l_k}=\mathcal{O}(10^{-4})$ even at $T=m_{\chi^3_{\pm}}$, exponentially decreases for $T \lesssim m_{\chi^3_{\pm}}$ (i.e. after $\chi^3_{\pm}$ becomes non-relativistic), thus leading to highly suppressed washout effects. We therefore observe that both $\chi^3_{\pm}$ decays are relevant for generating the lepton asymmetry.}
 Note that $\chi^3_{\pm}$ dominantly contribute to $n_B/s$, whereas $\chi^1_{\pm}, \chi^2_{\pm}$ give almost null contributions. Further, it is understood that the loops of the higgsino and leptons do not generate the lepton asymmetry because of $Im(\sum_{m=1}^3 y_{3_{+}}^{\prime m} y_{3_-}^{\prime m \ast} \sum_{n=1}^3 y_{3_+}^{\prime n \ast} y_{3_{-}}^{\prime n})=0$. %Im(y_{3_{+}}^{\prime n} y_{3_-}^{\prime n \ast} y_{3_+}^{\prime n \ast}y_{3_{-}}^{\prime n})=0$.} 
Eventually, we obtain $n_B/s \sim 8.7 \times 10^{-11}$ given that the off-diagonal element in Eq.(\ref{matrix-1}) is set as follows: 
\begin{equation} \left|\eta_{34} \eta^{\ast}_{43}+ \kappa^{\ast}_1 \kappa_2+ \kappa_1 \kappa^{\ast}_2 +7\left|\kappa_2 \right|^2\right|=0.6 \sqrt{2} %0.6+0.6i, 
\end{equation} 
which indicates $\phi_{cp}\sim \pi/4$. Regarding this evaluation, we emphasize that the sphaleron process is implemented because the decay rate of $\chi^3_{\pm}$ into the higgsino and the SM lepton is much larger than the Hubble parameter around $T=100$ GeV:
\begin{gather}
\dfrac{\Gamma_{\chi^{a}_{+} \rightarrow \tilde{h}_u,  l_n}}{H} \simeq 10^2 \times \left(\dfrac{100}{g_{\ast}\left(100\ \text{GeV}\right)}\right)^{1/2} \left(\dfrac{100\ \text{GeV}}{T}\right)^2,\\
\dfrac{\Gamma_{\chi^{a}_{-} \rightarrow \tilde{h}_u,  l_n}}{H} \simeq 2.34 \times 10^3 \times \left(\dfrac{100}{g_{\ast}\left(100\ \text{GeV}\right)}\right)^{1/2} \left(\dfrac{100\ \text{GeV}}{T}\right)^2.
\end{gather}

\section{Neutrino mass from See-Saw mechanism and $R$-parity violation}

\noindent
We estimate the neutrino mass. In our framework, some fermion components of $S^{ij}$ (denoted $\psi_{S^{ij}}$) provide sizable contributions through the See--Saw mechanism \cite{Yanagida:1979as}. 
$\psi_{S^{1i}}, \psi_{S^{i1}}\  (i=1-4)$ have the mass of $\sim \sqrt{m_1 \Lambda}$, while the mass of $\psi_{S^{ij}}\ (i,j=2-4)$ is solely induced from $b_i, \bar{b}_i, S^{ij}$ loops and the SUGRA correction \cite{Kitano:2006wz}. 
Among the effective $K\ddot{a}hler$ potential are the relevant couplings:
\begin{align}
K_{eff} =& \sum_{i < j} \dfrac{\left|S^{ii}\right|^2+\left|S^{jj} \right|^2}{48 \pi^2 m_1 \Lambda} \left(\left|S^{ij} \right|^2+ \left|S^{ji} \right|^2\right) \notag \\ &+ \sum_{i} \dfrac{\left|S^{ii}\right|^4}{96 \pi^2 m_1 \Lambda}, 
\end{align}
where $i, j$ runs $2-4$.
$S^{ii} \ (i=2-4)$ develops the VEV:
\begin{equation}
\left<S^{ii} \right>= 3 m_{3/2} \cdot \dfrac{48 \pi^2 m_1 \Lambda F_{ii}}{2 \left|F_{ii}\right|^2},
\end{equation}
via the following scalar potential:
\begin{align}
V \supset \dfrac{2 m_{i}^{2}\Lambda^2}{48 \pi^2 m_1\Lambda} \left|S^{ii}\right|^2-3 m_{3/2} F_{ii} S^{ii}+h.c.,
\end{align}
thus giving rise to the mass of $\psi_{S^{ij}}\ (i,j=2-4)$:
\begin{equation}
m_{\psi_{S^{ij}}} \simeq 3 m_{3/2}.
\end{equation}
The neutrino then acquires the mass of
\begin{align}
m_{\nu_n} \simeq \ \dfrac{\sum_{i,j=2}^4 \left|y_{ij}^n\right|^2 v^2 \sin^2{\beta}}{3 m_{3/2}} \left(\dfrac{\Lambda}{M_{pl}}\right)^2.
\end{align}
Here, $v=174$ GeV is the VEV of the Higgs boson and $\tan{\beta}$ is the ratio of $\left<H_u\right>$ to $\left<H_d\right>$. Further, it should be noted that the $R$-parity violation induces the VEV of the MSSM sneutrino, yielding extra contributions to the neutrino masses; among the scalar potential is the Higgs-slepton mixing term:\begin{align}
V \supset  \dfrac{\Lambda^2}{M_{pl}} & \left|\sum_{i=2}^4 y_{ii}^nm_i \right|  H_u \tilde{L}_n.  %\simeq 4.2 \times 10^6 \ \left[\text{GeV}\right]^2 \notag \\ & \times \left(\dfrac{\left|y_{22}^n +0.2y_{33}^n+0.1y_{44}^n \right|}{10^{-4}}\right) H_u \tilde{L}_n. 
\end{align}The MSSM  sneutrino (denoted $\tilde{\nu}_n$) thus develops the VEV:
\begin{equation}
\left<\tilde{\nu}_n\right> \simeq \dfrac{\Lambda^2}{M_{pl}}  \left|\sum_{i=2}^4 y_{ii}^nm_i \right| \cdot \dfrac{v\sin {\beta}}{m^2_{\tilde{\nu}_n}}.
\end{equation}
Consequently, the neutrino receives extra contributions in its mass matrix:
\begin{align}
\left(\Delta m_{\nu}\right)_{nm} \simeq & \ m^2_{Z^0} \dfrac{\left<\tilde{\nu}_n\right> \left<\tilde{\nu}_m\right>}{v^2 m_{\chi^0}}\notag \\ & \simeq 8.32 \times \left[1.05 \times 10^5 y^n_{22}+2.09 \times 10^4 y^n_{33}+1.05 \times 10^4 y^n_{44}\right] \notag \\ &\times \left[1.05 \times 10^5 y^m_{22}+2.09 \times 10^4 y^m_{33}+1.05 \times 10^4 y^m_{44}\right] \notag \\ & \times \left(\dfrac{\sin {\beta}}{1.0}\right)^2\left(\dfrac{1\ \text{TeV}}{m_{\tilde{\nu}_n}}\right)^4 \left(\dfrac{1\ \text{TeV}}{m_{\chi^0}}\right),
\end{align}
where $m_{Z^0}, m_{\chi^0}$ denote the $Z^0$ boson mass of $\sim 91.2$ GeV and the neutralino mass, respectively.
The cosmologically allowed mass of the neutrino implies $\left(\Delta m_{\nu}\right)_{nm} \lesssim 10^{-1}\ \text{eV}$, entailing a strict limitation on $y^n_{ii} (i=2-4)$: 
\begin{align}
&\left[1.05 \times 10^5 y^n_{22}+2.09 \times 10^4 y^n_{33}+1.05 \times 10^4 y^n_{44}\right] \notag \\ &\lesssim 4.95 \times 10^{-6} \cdot \left(\dfrac{0.7}{\sin {\beta}}\right)\left(\dfrac{m_{\tilde{\nu}_n}}{1\ \text{TeV}}\right)^2 \left(\dfrac{m_{\chi^0}}{1\ \text{TeV}}\right)^{1/2}.  \label{limitation}
\end{align}
Altogether, we evaluate $m_{\nu_n}$ as follows:
\begin{equation}
m_{\nu_n} \simeq 4.7 \times 10^{-1}\ \text{eV} \ \left(\dfrac{\sin {\beta}}{0.7} \right)^2 \left(\dfrac{\Lambda}{10^{11.7}\ \text{GeV}}\right)^2, 
\end{equation}
%\begin{equation}
%m_{\nu_n} \simeq 2.8 \times 10^{-1}\ \text{eV} \ \left(\dfrac{\sin {\beta}}{0.7} \right)^2 \left(\dfrac{\Lambda}{10^{11.7}\ \text{GeV}}\right)^2, 
%\end{equation}
setting the parameters of $\left|y_{34}^n \right|^2=\left|y_{43}^n \right|^2=0.2, \ \left|y_{23}^n \right|^2=\left|y_{32}^n \right|^2=\left|y_{24}^n \right|^2=\left|y_{42}^n \right|^2=10^{-1.6}$ and $\left|y_{ii}^n \right|<10^{-10}$.
\\

\section{Decay of dual singlet fermions}

\noindent
Let us address the decay of $\psi_{S^{ij}} \ (i,j=2-4)$ except for the gravitino (that consists of a linear combination of $\psi_{S^{ij}} (i=2-4)$). Noting that the very eight fermions slightly mix with the neutrino, one verifies that they promptly decay into the SM leptons or quarks via the $Z^0$ boson or the $W^{\pm}$ boson, respectively. The decay widths of relevance are as follows:
\begin{gather}
\Gamma_{\psi_{S^{ij}} \rightarrow \ 3 \nu} \simeq \dfrac{9 \zeta^2g_2^4}{3840 \pi^3\cos^4{\theta_W}}\cdot\dfrac{17}{16} \cdot \dfrac{m^5_{\psi_{S^{ij}}}}{m^4_{Z^0}}, \label{three-body-decay1}\\
\Gamma_{\psi_{S^{ij}} \rightarrow \ {\nu e^+ e^-}} \simeq \dfrac{3 \zeta^2g_2^4}{3840 \pi^3\cos^4{\theta_W}} \cdot\dfrac{17}{16}\cdot \dfrac{m^5_{\psi_{S^{ij}}}}{m^4_{Z^0}} C_e, \\
\Gamma_{\psi_{S^{ij}} \rightarrow \ {\nu \bar{u} u}} \simeq \dfrac{3 \zeta^2g_2^4}{3840 \pi^3\cos^4{\theta_W}}\cdot\dfrac{17}{16} \cdot \dfrac{m^5_{\psi_{S^{ij}}}}{m^4_{Z^0}} C_u, \\
\Gamma_{\psi_{S^{ij}} \rightarrow \ {\nu \bar{d} d}} \simeq \dfrac{3 \zeta^2g_2^4}{3840 \pi^3\cos^4{\theta_W}} \cdot\dfrac{17}{16}\cdot \dfrac{m^5_{\psi_{S^{ij}}}}{m^4_{Z^0}} C_d, \\
\Gamma_{\psi_{S^{ij}} \rightarrow \  e^-  e^+ \nu} \simeq \dfrac{4 \zeta^2g_2^4}{3840 \pi^3} \cdot\dfrac{17}{16}\cdot \dfrac{m^5_{\psi_{S^{ij}}}}{m^4_{W^+}}, \\
\Gamma_{\psi_{S^{ij}} \rightarrow \ e^+  u \bar{d}} \simeq \dfrac{4 \zeta^2g_2^4}{3840 \pi^3} \cdot\dfrac{17}{16}\cdot \dfrac{m^5_{\psi_{S^{ij}}}}{m^4_{W^+}}\cdot \cos^2{\theta_c}, \\
\Gamma_{\psi_{S^{ij}} \rightarrow \  e^+  e^- \nu} \simeq \dfrac{4 \zeta^2g_2^4}{3840 \pi^3} \cdot\dfrac{17}{16}\cdot \dfrac{m^5_{\psi_{S^{ij}}}}{m^4_{W^-}}, \\
\Gamma_{\psi_{S^{ij}} \rightarrow \ e^-  \bar{u} d} \simeq \dfrac{4 \zeta^2g_2^4}{3840 \pi^3} \cdot\dfrac{17}{16}\cdot \dfrac{m^5_{\psi_{S^{ij}}}}{m^4_{W^-}} \cdot \cos^2{\theta_c},
\\
\Gamma_{\psi_{S^{ij}} \rightarrow \  \mu^-  e^+ \nu} \simeq \dfrac{4 \zeta^2g_2^4}{64 \pi^3} D_+,\\
\Gamma_{\psi_{S^{ij}} \rightarrow \  \mu^-  \bar{u} d} \simeq \dfrac{4 \zeta^2g_2^4}{64 \pi^3} D_+\cdot \cos^2{\theta_c}, \\
\Gamma_{\psi_{S^{ij}} \rightarrow \  \mu^+  e^- \nu} \simeq \dfrac{4 \zeta^2g_2^4}{64 \pi^3} D_-,
\end{gather}
\begin{gather}
\Gamma_{\psi_{S^{ij}} \rightarrow \  \mu^+  u \bar{d}} \simeq \dfrac{4 \zeta^2g_2^4}{64 \pi^3} D_-\cdot \cos^2{\theta_c}, \\
\Gamma_{\psi_{S^{ij}} \rightarrow \  e^- \mu^+\nu} \simeq \dfrac{4 \zeta^2g_2^4}{64 \pi^3} D_+,\\
\Gamma_{\psi_{S^{ij}} \rightarrow \  e^+ \mu^-\nu} \simeq \dfrac{4 \zeta^2g_2^4}{64 \pi^3} D_+ \label{three-body-decay2}
\end{gather}
with $C_{e, u, d}, D_{\pm}$ given by
\begin{align}
&C_e \equiv \left(-\dfrac{1}{2}+\sin^2{\theta_W}\right)^2 + \sin^4{\theta_W},\\
&C_u \equiv \left(\dfrac{1}{2}-\dfrac{2}{3}\sin^2{\theta_W}\right)^2+ \dfrac{4}{9} \sin^4{\theta_W}, \\
&C_d \equiv \left(-\dfrac{1}{2}+\dfrac{1}{3}\sin^2{\theta_W}\right)^2+ \dfrac{1}{9} \sin^4{\theta_W}, 
%&D_+ \equiv \dfrac{1}{12} \dfrac{m^5_{\psi_{S^{ij}}}}{m^4_{W^+}}\left(1-\dfrac{m_{\mu^+}}{m_{\psi_{S^{ij}}}}\right)^4 \notag \\ &\ \ \ -\dfrac{1}{15} \dfrac{m^5_{\psi_{S^{ij}}}}{m^4_{W^+}}\left(1-\dfrac{m_{\mu^+}}{m_{\psi_{S^{ij}}}}\right)^5,
\end{align}
\begin{align}
&D_+ \equiv \dfrac{1}{24} \dfrac{m^5_{\psi_{S^{ij}}}}{m^4_{W^+}}\left(1-\dfrac{m^2_{\mu^+}}{m^2_{\psi_{S^{ij}}}}\right)^3 + \dfrac{1}{16} \dfrac{m^4_{\psi_{S^{ij}}} \cdot m_{\mu^+}}{m^4_{W^+}}\left(1-\dfrac{m^2_{\mu^+}}{m^2_{\psi_{S^{ij}}}}\right)^3\notag \\ &\ \ \ \ +\dfrac{1}{24}\dfrac{m^3_{\psi_{S^{ij}}} \cdot m^2_{\mu^+}}{m^4_{W^+}}\left(1-\dfrac{m^2_{\mu^+}}{m^2_{\psi_{S^{ij}}}}\right)^3-\dfrac{1}{192} \dfrac{m^5_{\psi_{S^{ij}}}}{m^4_{W^+}}\left(1-\dfrac{m^2_{\mu^+}}{m^2_{\psi_{S^{ij}}}}\right)^4 \notag \\ &\ \ \ \ -\dfrac{3}{160} \dfrac{m^5_{\psi_{S^{ij}}}}{m^4_{W^+}}\left(1-\dfrac{m^2_{\mu^+}}{m^2_{\psi_{S^{ij}}}}\right)^5, \label{D+}
\end{align}
\begin{align}
&D_- \equiv \dfrac{1}{24} \dfrac{m^5_{\psi_{S^{ij}}}}{m^4_{W^-}}\left(1-\dfrac{m^2_{\mu^-}}{m^2_{\psi_{S^{ij}}}}\right)^3 + \dfrac{1}{16} \dfrac{m^4_{\psi_{S^{ij}}} \cdot m_{\mu^-}}{m^4_{W^-}}\left(1-\dfrac{m^2_{\mu^-}}{m^2_{\psi_{S^{ij}}}}\right)^3\notag \\ &\ \ \ \ +\dfrac{1}{24}\dfrac{m^3_{\psi_{S^{ij}}} \cdot m^2_{\mu^-}}{m^4_{W^-}}\left(1-\dfrac{m^2_{\mu^-}}{m^2_{\psi_{S^{ij}}}}\right)^3-\dfrac{1}{192} \dfrac{m^5_{\psi_{S^{ij}}}}{m^4_{W^-}}\left(1-\dfrac{m^2_{\mu^-}}{m^2_{\psi_{S^{ij}}}}\right)^4 \notag \\ &\ \ \ \ -\dfrac{3}{160} \dfrac{m^5_{\psi_{S^{ij}}}}{m^4_{W^-}}\left(1-\dfrac{m^2_{\mu^-}}{m^2_{\psi_{S^{ij}}}}\right)^5, \label{D-}
\end{align}
%\begin{align}
%&D_- \equiv \dfrac{1}{12} \dfrac{m^5_{\psi_{S^{ij}}}}{m^4_{W^-}}\left(1-\dfrac{m_{\mu^-}}{m_{\psi_{S^{ij}}}}\right)^4 \notag \\ &\ \ -\dfrac{1}{15} \dfrac{m^5_{\psi_{S^{ij}}}}{m^4_{W^-}}\left(1-\dfrac{m_{\mu^-}}{m_{\psi_{S^{ij}}}}\right)^5,
%\end{align}
where $\theta_W, \theta_c$ are the Weinberg angle or the Cabibbo angle with $\sin^2{\theta_W} \simeq 0.23$ and $\tan{\theta_c} \simeq 0.23$ respectively, %the value of which approximately equals $\sin^2{\theta_W} \simeq 0.23$ %with $C_{e, u, d}$ totally adding up:
%\begin{equation}
%C_{total}=\dfrac{3}{4}-2 \sin^2{\theta_W}+\dfrac{28}{9}\sin^4{\theta_W},
%\end{equation}
 %and $\theta_c$ is the Cabibbo angle of $\tan{\theta_c} \simeq 0.23$, 
while $g_2$ denotes the $SU(2)_L$ gauge coupling constant.
Besides, $\zeta$ is defined as
\begin{equation}
\zeta \equiv \dfrac{y_n \Lambda}{M_{pl}} \cdot \dfrac{v \sin{\beta}}{m_{\psi_{S^{ij}}}},
\end{equation}
which represents the magnitude of mixing involved. 
The lifetime is then estimated as
\begin{gather}
\tau_{\psi_{S^{34}}} \simeq  \tau_{\psi_{S^{43}}} \simeq 3.97\times 10^{-2} \ \left(\dfrac{0.2}{\left|y_n\right|^2}\right) \left(\dfrac{10^{11.7}\ \text{GeV}}{\Lambda}\right)^2 \left(\dfrac{0.7}{\sin{\beta}}\right)^2 s,\\
\tau_{\psi_{S^{ij}}}^{OT}  \simeq 0.316\ \left(\dfrac{10^{-1.6}}{\left|y_n\right|^2}\right) \left(\dfrac{10^{11.7}\ \text{GeV}}{\Lambda}\right)^2 \left(\dfrac{0.7}{\sin{\beta}}\right)^2 s. \label{ot1}
\end{gather}
%\begin{gather}
%\tau_{\psi_{S^{34}}} \simeq  \tau_{\psi_{S^{43}}} \simeq 0.137\ \left(\dfrac{10^{-0.5}}{y_n}\right)^2 \left(\dfrac{10^{11.7}\ \text{GeV}}{\Lambda}\right)^2 \left(\dfrac{0.7}{\sin{\beta}}\right)^2 s,\\
%\tau_{\psi_{S^{ij}}}^{OT}  \simeq 0.545\ \left(\dfrac{10^{-0.8}}{y_n}\right)^2 \left(\dfrac{10^{11.7}\ \text{GeV}}{\Lambda}\right)^2 \left(\dfrac{0.7}{\sin{\beta}}\right)^2 s. \label{ot1}
%\end{gather}
This corresponds to the decay temperature of $\psi_{S^{ij}}$:
\begin{equation}
T^{decay}_{\psi_{S^{34}}} \simeq T^{decay}_{\psi_{S^{43}}} \simeq 5.02\ \text{MeV} \left(\dfrac{\left|y_n\right|^2}{0.2}\right)^{1/2} \left(\dfrac{\Lambda}{10^{11.7}\ \text{GeV}}\right) \left(\dfrac{\sin{\beta}}{0.7}\right),
\end{equation}
%\begin{equation}
%T^{decay}_{\psi_{S^{34}}} \simeq T^{decay}_{\psi_{S^{43}}} \simeq 2.69\ \text{MeV} \left(\dfrac{y_n}{10^{-0.5}}\right) \left(\dfrac{\Lambda}{10^{11.7}\ \text{GeV}}\right) \left(\dfrac{\sin{\beta}}{0.7}\right),
%\end{equation}
\begin{equation}
T_{\psi_{S^{ij}}}^{OT\ decay} \simeq 1.78\ \text{MeV} \left(\dfrac{\left|y_n\right|^2}{10^{-1.6}}\right)^{1/2} \left(\dfrac{\Lambda}{10^{11.7}\ \text{GeV}}\right) \left(\dfrac{\sin{\beta}}{0.7}\right), \label{ot2}
\end{equation}
%\begin{equation}
%T_{\psi_{S^{ij}}}^{OT\ decay} \simeq 1.35\ \text{MeV} \left(\dfrac{y_n}{10^{-0.8}}\right) \left(\dfrac{\Lambda}{10^{11.7}\ \text{GeV}}\right) \left(\dfrac{\sin{\beta}}{0.7}\right), \label{ot2}
%\end{equation}
%with $m_{\psi_{S^{ij}}}=372$ MeV, 
when their energy density accounts for only a small amount of the total energy density:
\begin{gather}
\dfrac{\rho_{{\psi_{S^{34}}}}+ \rho_{{\psi_{S^{43}}}}}{\rho_{rad.}} \simeq 0.122, \\ %0.118, 
\dfrac{\sum_{OT} \rho_{{\psi_{S^{ij}}}}}{\rho_{rad.}} \simeq 9.25 \times 10^{-2}. \label{ot3} %0.122.  
\end{gather}
Here, the superscript or subscript $OT$ in the left-hand side of Eqs. (\ref{ot1}), (\ref{ot2}) and (\ref{ot3}) denotes the other fermions, i.e. $\psi_{S^{ij}}$ for $(i, j)=(2, 3), (2, 4), (3, 2), (4, 2)$. (See Appendix C for the detailed analysis, based on the method in \cite{Plumacher:1997ru, Buchmuller:2004nz}.)
Eventually, we conclude that the presence/decay of these fermions only causes the universe to be reheated slightly, and thus has less significant implications on cosmology. 
\\

\section{Gravitino problem ameliorated under $R$-parity induced neutrino mass}

\noindent
Our model assumes the $Z_{4 R}$ violating (and therefore, the $R$-parity violating) interaction. This implies that the gravitino decays into lighter particles. 
The goldstino (denoted $\tilde{G}$) is written as a linear combination of $\psi_{S^{ii}} \ (i=2-4)$: 
\begin{equation}
\tilde{G} = \dfrac{1}{F_{total}} \sum_{i=2}^4 F_{S^{ii}} \psi_{S^{ii}}.
\end{equation}
We then discuss the gravitino decay width through its mixing with the neutrino, similar to the previous analysis. Defining $\xi$ (replacing by $\zeta$) as follows:
\begin{equation}
\xi \equiv \dfrac{\Lambda}{M_{pl}} \dfrac{1}{F_{total}} \left(\sum_{i=2}^4 y_{ii}^n F_{S^{ii}} \right) \left(\dfrac{v \sin \beta}{m_{3/2}}\right)
\end{equation} 
and considering Eq.(\ref{limitation}) with $m_{\psi_{S^{ij}}}$ replaced by $m_{3/2}$, we obtain the decay rate:
\begin{equation}
\Gamma_{\psi^{\mu} \rightarrow SM fermions} \lesssim 2.49 \times 10^{-42}\ \text{MeV}, \label{dominant-decay}
\end{equation}
%we obtain the lifetime:
%\begin{equation}
%\tau_{3/2} \gtrsim 2.64 \times 10^{20}\ s.
%\end{equation}
where ''SM fermions'' stands for the decay products shown in Eqs.(\ref{three-body-decay1})$-$(\ref{three-body-decay2}). Also, we omitted the contributions from Eqs.(\ref{D+}) and (\ref{D-}) because both yield less considerable effects. Meanwhile, Eq.(\ref{dual-yukawa}) leads to the interaction of
\begin{align}
\mathcal{L} \supset \dfrac{1}{F_{total}} \left(\dfrac{\Lambda}{M_{pl}} \right)  \sum_{n=1}^3 \sum_{i=2}^4 y_{ii}^n F_{S^{ii}} H_u  \tilde{G} \ l_n+h.c.. \label{goldstino-int-1}
\end{align}
Our model provides $m_{3/2} \sim 124$ MeV $> 2m_e+m_{\nu_e}$, and among the SM is the interaction:
\begin{equation} 
\mathcal{L} \supset Y_e H_d \ e^{+} e^- \supset Y_e  \cos{\beta} H^0 e^{+} e^-,
\end{equation}
where $Y_e$ is the Yukawa coupling constant, expressed as $Y_e= m_e/ v \cos{\beta}$ with $m_e \sim 0.51$ MeV denoting the electron mass, and $H^0$ is the lightest Higgs boson with its mass of $\sim 125$ GeV \cite{ATLAS:2012yve, CMS:2012qbp}. The gravitino thus decays through the effective Lagrangian density:
\begin{align}
\mathcal{L}_{eff} =  & \dfrac{\sqrt{3}}{F_{total}} \left(\sum_{i=2}^4 y_{ii}^1 F_{S^{ii}} \right)  \cdot \dfrac{\Lambda}{M_{pl}} \cdot \dfrac{\sin{\beta}}{m^2_{H^0}} \notag \\ & \times \dfrac{m_e}{v}\ \gamma_{\mu} \psi^{\mu}\  \nu_e e^{+} e^{-} + h.c..
\end{align}
The decay rate should be roughly estimated as follows \cite{three-body-decay}:
\begin{align}
\Gamma_{\psi^{\mu}\rightarrow  \nu_e e^{+} e^{-}} \simeq  \dfrac{\left|\sum_{i=2}^4 y_{ii}^1 F_{S^{ii}}\right|^2}{1280 \pi^3 F^2_{total}} \left(\dfrac{\Lambda}{M_{pl}}\right)^2  \cdot \left(\dfrac{\sin^2{\beta}}{m^4_{H^0}}\right) \cdot \left(\dfrac{m_e}{v}\right)^2 \cdot m^5_{3/2} < \mathcal{O}(10^{-61}) \ \text{MeV}. 
\end{align}
%The lifetime is then given by \cite{three-body-decay}
%\begin{align}
%\tau_{3/2} &\gtrsim 1.47 \times 10^{19} \notag \\ & \left(0.98 \left|y_{22}^1\right|+0.19 \left|y_{33}^1\right| +9.8 \times 10^{-2} \left|y_{44}^1\right|\right)^{-2}   s.
%\end{align} 
Additionally, the gravitino could decay into a photon and a neutrino:
\begin{equation}
\Gamma_{\psi^{\mu}\rightarrow \gamma \nu_n} \simeq \dfrac{1}{32 \pi} \cdot \dfrac{m_{\nu_{n}} \cos^2{\theta_{W}}}{m_{\chi^0}} \cdot \dfrac{m^3_{3/2}}{M_{pl}^2}=\mathcal{O}(10^{-51}) \ \text{MeV}.
\end{equation}
To summarize, $\Gamma_{\psi^{\mu} \rightarrow SM fermions}$ represents the predominant decay rate, corresponding to the gravitino lifetime of
\begin{equation}
\tau_{3/2} \gtrsim 2.64 \times 10^{20}\ s,
\end{equation}
which is much longer than the age of the universe. Furthermore, we stress that the long-lived gravitino leaves its abundance within the observed dark matter density of $\Omega_{DM} h^2 \sim 0.12$ depending on the gluino mass \cite{Moroi:1993mb, deGouvea:1997afu, Hook:2018sai}:
\begin{equation}
T_{RH} \lesssim 3.28\times10^4 \  \text{GeV}  \left(\dfrac{7.0\ \text{TeV}}{M_{gluino}}\right)^2 \left(\dfrac{m_{3/2}}{124\ \text{MeV}}\right).
\end{equation}
%To summarize, the gravitino could possess a longer lifetime than the age of the universe.

\section{Cosmic spectra from gravitino decay}

\noindent
Let us briefly address the effect of the unstable (even though long-lived) gravitino on observations of cosmic ray spectra. Here, we focus on the gamma-ray and the neutrino to estimate the magnitude of both fluxes. 
First, the gamma-ray flux from the two-body decay is evaluated. The extragalactic flux indicates a monochromatic line of
%\begin{equation}
%E^2 \dfrac{dJ^{\gamma}_{eg}}{dE} = \dfrac{\Omega_{3/2} \rho_c}{4 \pi \tau_{3/2} H_0 \Omega_M} \left[1+\kappa \left(\dfrac{2E}{m_{3/2}}\right)^{3}\right]^{-1/2}\left(\dfrac{2E}{m_{3/2}}\right)^{5/2} \theta \left(1-\dfrac{2E}{m_{3/2}}\right)
%\end{equation}
\begin{equation}
E^2 \dfrac{dJ^{\gamma}_{eg}}{dE} = \dfrac{\Omega_{3/2} \rho_c \cdot BR\left(\tilde{G} \rightarrow \gamma  \nu \right)}{8 \pi \tau_{3/2} H_0 \Omega_M} \left(1+\kappa x^{3}\right)^{-1/2}x^{5/2} \times \theta \left(1-x\right),
\end{equation}
where $\theta \left(1-x\right)$ is the Heaviside step function. The peak appears at $x=1$ as follows:
\begin{gather}
E^2 \dfrac{dJ^{\gamma}_{eg}}{dE} \simeq 9.58 \times 10^{-10}
 \left(\dfrac{1.0\ \text{TeV}}{m_{\chi^0}}\right) \left(\dfrac{m_{\nu_n}}{4.7 \times 10^{-1}\ \text{eV}}\right)\  \text{GeV$ \ s^{-1} str^{-1} cm^{-2}$}.
\end{gather}
Meanwhile, an additional flux comes from the dark matter halo:
\begin{gather}
E^2 \dfrac{dJ^{\gamma}_{halo}}{dE} = \dfrac{BR\left(\tilde{G}\rightarrow \gamma \nu\right)}{8 \pi \tau_{3/2}} \times \mathcal{I}_{halo}  \cdot \delta\left(1-\dfrac{2E}{m_{3/2}}\right), \label{halo-gamma-1}
\end{gather}
with 
\begin{equation}
\mathcal{I}_{halo} \equiv \left[\ \int_{l.o.s.}  \rho_{halo}\left(\vec{l}\right) \cdot d \vec{l}\  \right].
\end{equation}
Following \cite{Covi:2008jy}, let us apply the Navarro, Frenk, and White (NFW) profile \cite{Navarro:1995iw}:
\begin{equation}
\rho_{halo}\left(\vec{l}\right) = \dfrac{\rho_0}{(r/r_c) \left[1+(r/r_c)\right]^2}\ ,
\end{equation}
where $\rho_0=0.26$ GeV/cm$^3$ and $r_c=20$ kpc. Considering that the distance from the Galactic center to the Sun %(denoted $r_{\odot}$) 
is around $8.5$ kpc, while the radius of the Galactic disc %(denoted $r_D$)
 is $\sim 2$ kpc, we obtain the numerical result of
\begin{gather}
\mathcal{I}_{halo} \simeq 1.51 \times 10^{23} \ \text{GeV} \ cm^{-2}.
\end{gather}
Eventually, Eq.(\ref{halo-gamma-1}) is expressed as
\begin{gather}
E^2 \dfrac{dJ^{\gamma}_{halo}}{dE} =1.09 \times 10^{-8} \left(\dfrac{1.0\ \text{TeV}}{m_{\chi^0}}\right) \left(\dfrac{m_{\nu_n}}{4.7 \times 10^{-1}\ \text{eV}}\right) \cdot \delta\left(1-\dfrac{2E}{m_{3/2}}\right)\ \text{GeV$ \ s^{-1} str^{-1} cm^{-2}$}.
\end{gather}
Thus, even taking account of the gamma-ray flux from the dark matter halo, the extragalactic gamma-ray flux from the gravitino decay is much less than the upper limitation extracted from the Energetic Gamma Ray Experiment Telescope (EGRET) data \cite{EGRET:1997qcq}
that is given by
\begin{gather}
E^2 \dfrac{dJ^{\gamma}_{}}{dE} \simeq 1.81 \times 10^{-6} \left(\dfrac{E}{\text{62\ MeV}}\right)^{-0.1} \text{GeV$ \ s^{-1} str^{-1} cm^{-2}$}.
\end{gather}
Subsequently, let us evaluate the implication of the neutrino emission due to the three-body decay. We derive the extragalactic flux of the neutrino as 
\begin{align}
E^3 \dfrac{dJ^{\nu_i}_{eg}}{dE} = & \dfrac{\Omega_{3/2} \rho_c \cdot \left[\sum BR\left(\tilde{G} \rightarrow \nu_i \bar{l} l\right)+ \sum BR\left(\tilde{G} \rightarrow \nu_i \bar{q} q\right)\right]}{4 \pi \tau_{3/2} H_0 \Omega_M} \notag \\ & \times \dfrac{240}{17} \cdot \dfrac{m_{3/2}}{2} \cdot x^3 \int_{1}^{\frac{1}{x}} dy \left(\dfrac{1}{48} x^4 y^4-\dfrac{1}{8} x^3 y^3+\dfrac{3}{16} x^2 y^2\right) \times \dfrac{1}{\sqrt{\kappa+y^3}},
\end{align}
where $\kappa = \Omega_{\Lambda}/\Omega_m \simeq 3$ and $x\equiv 2E/m_{3/2}, \ y\equiv 1+z$. Besides, it is understood that the upper limit of integration interval, i.e. $1/x$ denotes the maximally allowed value of $y_{max} = m_{3/2}/2E$. Meanwhile, the flux from the dark matter halo takes the form of
\begin{align}
E^3 \dfrac{dJ^{\nu_i}_{halo}}{dE}  = \dfrac{\left[\sum BR\left(\tilde{G} \rightarrow \nu_i \bar{l} l\right)+ \sum BR\left(\tilde{G} \rightarrow \nu_i \bar{q} q\right)\right]}{4 \pi \tau_{3/2}} \times \mathcal{I}_{halo} \cdot \dfrac{dN_{\nu_{i}}}{dE},  
\end{align}
where $dN_{\nu_{i}}/dE$ is given by
\begin{equation}
\dfrac{dN_{\nu_{i}}}{dE}=\dfrac{240}{17} \cdot \dfrac{m_{3/2}}{2} \left(\dfrac{1}{48} x^7-\dfrac{1}{8} x^6+\dfrac{3}{16} x^5\right).
\end{equation}
We thus estimate the maximum of each neutrino flux, corresponding to $x=1$, as follows:
%\begin{gather}
%E^3 \dfrac{dJ^{\nu_e}_{eg}}{dE} \simeq 1.01 \times 10^{-3} \left(\dfrac{10^{-11}}{y^n_{22}+ 0.199 \times y^n_{33}+0.1 \times y^n_{44}}\right)^2 \left(\dfrac{0.7}{\sin \beta}\right)^2 \ \text{GeV$^2 s^{-1} str^{-1} cm^{-2}$}, \\
%E^3 \dfrac{dJ^{\nu_{\mu}}_{eg}}{dE} \simeq E^3 \dfrac{dJ^{\nu_{\tau}}_{eg}}{dE} \simeq 4.86 \times 10^{-4} \left(\dfrac{10^{-11}}{y^n_{22}+ 0.199 \times y^n_{33}+0.1 \times y^n_{44}}\right)^2 \left(\dfrac{0.7}{\sin \beta}\right)^2\ \text{GeV$^2 s^{-1} str^{-1} cm^{-2}$},
%\end{gather} 
\begin{gather}
E^3 \dfrac{dJ^{\nu_e}_{total}}{dE} \simeq 6.09 \times 10^{-2} \left(\dfrac{10^{-11}}{y^n_{22}+ 0.199 \times y^n_{33}+0.1 \times y^n_{44}}\right)^2 \left(\dfrac{0.7}{\sin \beta}\right)^2 \ \text{GeV$^2 s^{-1} str^{-1} cm^{-2}$}, \\ %nu-e flux 
E^3 \dfrac{dJ^{\nu_{\mu}}_{total}}{dE} \simeq 2.94 \times 10^{-2} \left(\dfrac{10^{-11}}{y^n_{22}+ 0.199 \times y^n_{33}+0.1 \times y^n_{44}}\right)^2 \left(\dfrac{0.7}{\sin \beta}\right)^2 \ \text{GeV$^2 s^{-1} str^{-1} cm^{-2}$}, \\
E^3 \dfrac{dJ^{\nu_{\tau}}_{total}}{dE} \simeq 2.94 \times 10^{-2} \left(\dfrac{10^{-11}}{y^n_{22}+ 0.199 \times y^n_{33}+0.1 \times y^n_{44}}\right)^2 \left(\dfrac{0.7}{\sin \beta}\right)^2 \ \text{GeV$^2 s^{-1} str^{-1} cm^{-2}$},
\end{gather}
setting $\Omega_{3/2}=0.1$. Altogether, we point out that the neutrino flux may exhibit a certain excess at $E = \mathcal{O}(10)$ MeV, while the gamma-ray line remains observationally acceptable. Particularly, the $\nu_e$ flux may be observed on the Earth  
\begin{equation}
\left[E^3 \dfrac{dJ^{\nu_e}_{total}}{dE}\right]_{Earth} \simeq 3.41 \times 10^{-2} \left(\dfrac{10^{-11}}{y^n_{22}+ 0.199 \times y^n_{33}+0.1 \times y^n_{44}}\right)^2 \left(\dfrac{0.7}{\sin \beta}\right)^2 \ \text{GeV$^2 s^{-1} str^{-1} cm^{-2}$}, \\ %nu-e flux
\end{equation}
via the neutrino oscillations, given the relevant probability of $P\left(\nu_e \rightarrow \nu_e\right)=0.56$ \cite{Grefe:2011dp}. 

\section{Pseudo NGB relics}

\noindent
We roughly discuss the implications of $\phi$ on cosmology. 
All the symmetries allow the following term among the superpotential:
\begin{align}
W  \supset c_1  \dfrac{\Lambda^4}{M_{pl}^3} b^2_1+c_2 \dfrac{\Lambda^4}{M_{pl}^3}\bar{b}^2_1 \ \supset c \dfrac{\Lambda^4}{M_{pl}^3} \dfrac{\left(b_1-\bar{b}_1\right)^2}{2},
\end{align} 
which makes $\phi$ massive:
\begin{align}
m_{\phi} &\simeq   \dfrac{\left|c \right| \Lambda^4}{M_{pl}^3} \notag \\ &\simeq 2.27 \ \text{eV} \left(\dfrac{\left|c \right|}{0.5}\right) \times \left(\dfrac{\Lambda}{10^{11.7}\ \text{GeV}}\right)^4,
\end{align}
where we set $c_1=c_2\equiv c$ for simplicity. Consequently, $\phi$ has the initial energy density of
\begin{align}
\rho^i_{\phi}& \simeq  m_{\phi}^2 m_1 \Lambda  \simeq 1.3 \times 10^5 \ \left[\text{GeV}\right]^4 \notag \\ & \times  \left(\dfrac{\left|c \right|}{0.5}\right)^2  \left(\dfrac{m_1}{10^{10.7}\ \text{GeV}}\right)\left(\dfrac{\Lambda}{10^{11.7}\ \text{GeV}}\right)^9.
\end{align}
The coherent oscillation begins around the temperature of
\begin{align}
T_{\phi} &\simeq \sqrt{m_{\phi}M_{pl}} \notag \\ &\simeq 7.46\times 10^4 \ \text{GeV} \cdot \left(\dfrac{m_{\phi}}{2.27\ \text{eV}}\right)^{1/2}. 
\end{align}
Taking into consideration that \cite{scale-factor} 
\begin{equation}
\rho_{\phi} \left(T \right) \simeq \rho^i_{\phi} \cdot \left(\dfrac{T_{RH}}{T_{\phi}}\right)^8 \left(\dfrac{T}{T_{RH}}\right)^3,
\end{equation}
we thus conclude that $\rho_{\phi}$ remains only a fraction of the matter energy density (denoted $\rho_{M}$) after the matter-dominated epoch begins:
\begin{align}
&\dfrac{\rho_{\phi} \left(T \lesssim T_{eq} \right)}{\rho_{M} \left(T \lesssim T_{eq} \right)} \simeq \dfrac{\rho_{\phi} \left(T_{eq} \right)}{\rho_{R} \left(T_{eq} \right)}\notag \\ \ \ \ \ \ \simeq &\  7.57 \times 10^{-2}  \left(\dfrac{\left|c \right|}{0.5}\right)^{1/2} \left(\dfrac{3.9}{g_{\ast}\left(T_{eq} \right)} \right) \notag \\ &\times \left(\dfrac{m_1}{10^{10.7}\ \text{GeV}}\right)  \left(\dfrac{\Lambda}{10^{11.7}\ \text{GeV}}\right)^3,
\end{align}
where $\rho_{R}$ is the radiation energy density:
\begin{equation}
\rho_{R} \left(T \right)=\dfrac{\pi^2 g_{\ast}\left(T \right)}{30} T^4,
\end{equation}
and $T_{eq} \sim 8.2 \times 10^{-10}$ GeV denotes the temperature that satisfies the relation:
\begin{equation}
\rho_{R} \left(T \right) = \rho_{M} \left(T \right).
\end{equation}

\section{Discussion}

\noindent 
As already evaluated, the $R$-parity violating interaction of $\psi_{S^{ii}}\ (i=2-4)$ inevitably involves a highly fine-tuning on $y^{ii}_n$ of $\mathcal{O}(10^{-11})$, though we may use an additional discrete symmetry, e.g. a $Z_4$ symmetry to provide a plausible prescription. The charge is assigned as shown in Table $3$:
\begin{center}
\begin{tabular}{c|c|c|c|c}
\multicolumn{4}{c}{Table 3. $Z_{4}$ Charge assignment}
\\ \toprule \addlinespace[2pt]
&    $Q^1, Q^2, Q^3, Q^4$ & $\bar{Q}^1, \bar{Q}^2, \bar{Q}^3, \bar{Q}^4$ & $Tr W^{\alpha}W_{\dot{\alpha}}$ & $H_u L_n$\\ \hline
\addlinespace[2pt] $Z_{4}$ & $2,\ 0,$\   $2,\ 0$ & $0,\ 2,$ \  $0,\ 2$ & $2$ & $0$\\ \bottomrule
\end{tabular}
\end{center}
Under such a circumstance, the $R$-violating interaction is reduced to the form of
\begin{align}
W^{RV}_{dual}\supset & \sum_{n=1}^3  \dfrac{\Lambda}{M_{pl}} \left(y_{34}^n S^{34}+y_{43}^n S^{43}\right) H_u L_n \label{crucial-term} \\ &+\sum_{n=1}^3 \ \sum_{i=2}^4 y_{ii}^n\dfrac{\Lambda \left<Tr W^{\alpha}W_{\dot{\alpha}}\right> S^{ij} H_u L_n}{M_{pl}^4}, \label{dual-yukawa-2}
\end{align}
which drastically relaxes the fine-tuning as follows:
\begin{equation}
y_{ii}^n \lesssim 10^{-3.7},
\end{equation}
while the thermal leptogenesis is implemented successfully, owing to the presence of the terms in Eq.(\ref{crucial-term}).\\

\section*{Conclusion}

\noindent
In this letter, we proposed the possibility that the dual singlet bosons, present in the ISS model, serve a crucial role in implementing thermal leptogenesis at low reheating temperature. Therein, we supposed that the hidden quarks have somewhat hierarchical masses, and we pointed out that the lepton asymmetry emerges from the pseudo-NGB loop and the quasi-degenerate mass among those dual singlet bosons. We thus constructed an explicit model that accommodates $T_{RH}\simeq 3 \times10^{4}$ GeV and the neutrino masses at $\mathcal{O}(0.1)$ eV, without any relation to the GUT breaking scale. Besides, we conducted analyses for the cosmological implications of the gravitino and the pseudo-NGB.  In our model, the $Z_{4 R}$ symmetry (and thus the $R$-parity) violation triggers the gravitino decay, though its lifetime can be longer than the age of the universe. Under such a circumstance, we verified that the gravitino abundance is within the observed DM density as long as the gluino mass is smaller than $\sim 7.0$ TeV. Regarding, we evaluated the effects of the long-lived gravitino on the spectra of cosmic rays. It is emphasized that the extragalactic neutrinos are emitted solely from the three-body decay in our model, which makes it possible for the neutrino flux to have a considerably large peak while the extragalactic gamma-ray flux is highly suppressed. Unfortunately, the observation in Super Kamiokande has so far suggested that the magnitude of the neutrino flux is $E^3 (dJ^{\nu_i}_{eg}/dE) = \mathcal{O}(10^{-4}-10^{-3})$ GeV $s^{-1} str^{-1} cm^{-2}$ at $E \sim 100$ MeV \cite{Honda:2015fha}. We thus expect that there may appear a clear excess of the neutrino (particularly, $\nu_e$) flux at forthcoming experiments such as Hyper Kamiokande.
 Further, we showed that if the pseudo-NGB is non-thermally produced, its relics only exhibit a less significant contribution.
 Finally, we proposed a discrete symmetry to render the neutrino mass within $\mathcal{O}\left(0.1\right)$ eV. As for this issue, an extended DSB sector, possessing a large flavor/ gauge symmetry as suggested by \cite{dual}, possibly realizes a certain Yukawa coupling structure in a compatible way with the cosmological observation of the neutrino mass. 

%Finally, we simply assumed the suppressed coupling of $S^{ii} \ (i=2-4)$ to $H_u L_i$ so that the SM slepton, whose soft mass is a few TeV, can be non-tachyonic. Meanwhile, with a more naive choice, i.e. $\left|y^n_{ii}\right|$ of $\sim 10^{-2} -10^{-1}$, their mass would be $\mathcal{O}(10)$ TeV, which is larger than the gluino mass by an order.
%Such a spectrum is seemingly realized in the sort of the gauge-mediation where the messengers carry the same quantum number as the Higgs belonging to $5+ \bar{5}$ of $SU(5)_{GUT}$ \cite{Raby:1997bpa}. Or, an alternative approach might be described in the language of the direct-mediated SUSY breaking model that possesses a large flavor/gauge symmetry \cite{Affleck:1984xz, Kitano:2006xg}.

{\section* {Acknowledgment}}

\noindent
I would like to thank S. Biondini (Basel University) for informing me of their work.
\\

\newpage
\begin{center}
{\bf \large Appendix}
\end{center}
\begin{center}
{\bf A.\  Loop-induced mass of $S^{ij}$ boson and fermion}
\end{center}
\noindent
We show that the $S^{ij} \ (i, j=2-4)$ bosons acquire the mass induced by the loop effect.  
Let us set $S^{ij}, b_i, \bar{b}_i$ for the $SU(3)$ gauge group with $N_F=4$:
\begin{gather}
S^{ij}= \begin{pmatrix}
\mathcal{X} & \mathcal{Y}^{l} \\
\mathcal{Z}^k & S_0^{kl} + \mathcal{S}^{kl}
\end{pmatrix} \notag \\
b_{i} = \begin{pmatrix}
\sqrt{m_1 \Lambda} \cdot \exp{\left(\phi_1 \right)}+ \zeta \\
\xi_{k}
\end{pmatrix}, \ \ \ 
\bar{b}_{i} = \begin{pmatrix}
\sqrt{m_1 \Lambda} \cdot  \exp{\left(-\phi_1 \right)} + \bar{\zeta} \\
\bar{\xi}_{k}
\end{pmatrix} \notag \\
 k, l=2-4. \tag{A.1}
\end{gather} 
Here, $S_0^{kl}$ and $\phi_{1}$ are the background fields.  In contrast, the others denote dynamical superfields corresponding to the fluctuations around
\begin{equation}
\left<S^{kl}\right>= {\bf{0_{\ 3 \times 3}}},\ \ \left<b_{1}\right>=\left<\bar{b}_{1}\right>=\sqrt{m_1 \Lambda},\ \  \left<b_{k}\right>=\left<\bar{b}_{k}\right>=0. \tag{A.2}
\end{equation}
  We then rewrite the relevant superpotential of Eq.(4) in the text as
\begin{align}
W_{dual}\supset \sqrt{m_1 \Lambda}\cdot \exp{\left(\phi_1 \right)} \mathcal{Y} \bar{\xi} + \sqrt{m_1 \Lambda}\cdot \exp{\left(-\phi_1 \right)} \mathcal{Z} \xi + \xi^{\mathrm{T}} \left(S_0^{kl} + \mathcal{S}^{kl}\right) \bar{\xi} -m_k \Lambda \left(S_0^{kk} + \mathcal{S}^{kk}\right). \tag{A.3}
\end{align}
(Notice that the other coupling of
\begin{align}
W_{dual} \supset \sqrt{m_1 \Lambda}\cdot \exp{\left(\phi_1 \right)}\mathcal{X}\bar{\zeta}+\sqrt{m_1 \Lambda}\cdot \exp{\left(-\phi_1 \right)}\mathcal{X}^{\mathrm{T}}\zeta \tag{A.4}
\end{align}
gives no contribution to the mass of the ${S}_0^{kl}$ boson.) We thus verify that ${S}_0^{kl}$ could acquire the loop--induced mass. It is then convenient to set $\phi_1=0$, and  
 the corrected scalar potential is given by 
\begin{equation}
V_{1 loop} = \dfrac{M_{B}^4}{64 \pi^2} \log{\left(\dfrac{M_{B}^2}{\Lambda^2_{CUT}}\right)}-\dfrac{M_{F}^4}{64 \pi^2} \log{\left(\dfrac{M_{F}^2}{\Lambda^2_{CUT}}\right)}=\dfrac{1}{4} Tr \int \dfrac{d^4 p}{\left(2\pi\right)^4} \dfrac{M_B^2}{p^2+M_B^2}-\dfrac{1}{4} Tr \int \dfrac{d^4 p}{\left(2\pi\right)^4} \dfrac{M_F^2}{p^2+M_F^2}. \tag{A.5}
\end{equation}
Here, $M_{B, F}$ is the boson or fermion mass matrix:
\begin{gather}
M^2_B= \begin{pmatrix} \mathcal{M}^{\dagger} \mathcal{M} & O \\ O & \mathcal{M}\mathcal{M}^{\dagger} \end{pmatrix}+ H, \ \ M^2_F= \begin{pmatrix} \mathcal{M}^{\dagger} \mathcal{M} & O \\ O & \mathcal{M}\mathcal{M}^{\dagger} \end{pmatrix}. \tag{A.6}
\end{gather}
$\mathcal{M}, H$ are written as
\begin{equation} \mathcal{M} = \begin{pmatrix} O & M \\ M & T \end{pmatrix},\ \ H=\begin{pmatrix} O & \mathcal{F}^{\dagger} \\ \mathcal{F} & O \end{pmatrix}, \tag{A.7} \end{equation}
where 
\begin{equation}
M = \begin{pmatrix} 0 & \sqrt{m_1 \Lambda}\cdot \bf{1}_{3 \times 3}\\  \sqrt{m_1 \Lambda} \cdot \bf{1}_{3 \times 3} & 0 \\  \end{pmatrix}, \ \ \ T= \begin{pmatrix} 0 & S_0 \\ S_0^{T} & 0 \end{pmatrix}, \ \ \mathcal{F}= \begin{pmatrix} O & O \\ O & F_M \end{pmatrix},\ \  F_M= \begin{pmatrix} 0 & F \\ F & 0 \end{pmatrix}, \tag{A.8}\end{equation}
and $F$ is given by
\begin{equation}
F= \begin{pmatrix} m_2 \Lambda & 0& 0 \\ 0 & m_3 \Lambda & 0  \\ 0 & 0 & m_4 \Lambda \end{pmatrix}, \tag{A.9}\end{equation}
whereas  $\Lambda_{CUT}$ is some cut-off scale. Besides, the row and column of $\mathcal{M}, \mathcal{F}$ run $\mathcal{Y}^{l}, \mathcal{Z}^k, \xi_{k}, \ \bar{\xi}_{k}$ respectively.
Rewriting Eq.(A.5) as
\begin{align}
V_{1 loop} =&\dfrac{1}{4} Tr \left[\left(m_1 \Lambda \cdot {\bf{1}_{24 \times 24}} + E^{\prime}+H\right) \int \dfrac{d^4 p}{\left(2\pi\right)^4} \dfrac{1}{p^2+ m_1 \Lambda} \left({\bf{1}_{24 \times 24}} + \dfrac{E^{\prime}+H}{p^2+ m_1 \Lambda}\right)^{-1}\right] \notag \\ 
& -\dfrac{1}{4} Tr \left[\left(m_1 \Lambda \cdot {\bf{1}_{24 \times 24}} + E^{\prime}\right) \int \dfrac{d^4 p}{\left(2\pi\right)^4} \dfrac{1}{p^2+ m_1 \Lambda} \left({\bf{1}_{24 \times 24}} + \dfrac{E^{\prime}}{p^2+ m_1 \Lambda}\right)^{-1}\right], \tag{A.10}
\end{align}
 one obtains the $S^{ij}\ (i,j=2-4)$ boson mass from the $1$--loop corrected scalar potential as follows:
\begin{gather}
\left(m^{loop}_{S^{23}}\right)^2 \equiv  \dfrac{\left(m_{2}^{2} + m_{3}^{2} \right)\Lambda^2}{48 \pi^2 m_1\Lambda} \left(\left|S^{23}\right|^2 +\left|S^{32}\right|^2\right), \tag{A.11} \\
\left(m^{loop}_{S^{24}}\right)^2 \equiv  \dfrac{\left(m_{2}^{2} + m_{4}^{2} \right)\Lambda^2}{48 \pi^2 m_1\Lambda} \left(\left|S^{24}\right|^2 +\left|S^{42}\right|^2\right), \tag{A.12} \\ 
\left(m^{loop}_{S^{34}}\right)^2 \equiv  \dfrac{\left(m_{3}^{2} + m_{4}^{2} \right)\Lambda^2}{48 \pi^2 m_1\Lambda} \left(\left|S^{34}\right|^2 +\left|S^{43}\right|^2\right), \tag{A.13} \\
\left(m^{loop}_{S^{22}}\right)^2 \equiv  \dfrac{2m_{2}^{2}\Lambda^2}{48 \pi^2 m_1\Lambda} \left|S^{22}\right|^2, \tag{A.14} \\ 
\left(m^{loop}_{S^{33}}\right)^2 \equiv  \dfrac{2m_{3}^{2}\Lambda^2}{48 \pi^2 m_1\Lambda} \left|S^{33}\right|^2, \tag{A.15} \\ 
\left(m^{loop}_{S^{44}}\right)^2 \equiv  \dfrac{2m_{4}^{2}\Lambda^2}{48 \pi^2 m_1\Lambda} \left|S^{44}\right|^2, \tag{A.16}
\end{gather}
with higher-order contributions:
\begin{gather} 
\left[\dfrac{\left(m_{2}^{4} + m_{3}^{4} \right) \Lambda^4}{1920 \pi^2 m_1^3 \Lambda^3}+\mathcal{O}\left(\dfrac{\left(m_{2}^{6} + m_{3}^{6} \right) \Lambda^6}{m_1^5 \Lambda^5}\right)\right] \left(\left|S^{23}\right|^2 +\left|S^{32}\right|^2\right), \tag{A.17} \\ \left[\dfrac{\left(m_{2}^{4} + m_{4}^{4} \right) \Lambda^4}{1920 \pi^2 m_1^3 \Lambda^3}+\mathcal{O}\left(\dfrac{\left(m_{2}^{6} + m_{4}^{6} \right) \Lambda^6}{m_1^5 \Lambda^5}\right)\right] \left(\left|S^{24}\right|^2 +\left|S^{42}\right|^2\right),  \tag{A.18}\\ \left[\dfrac{\left(m_{3}^{4} + m_{4}^{4} \right) \Lambda^4}{1920 \pi^2 m_1^3 \Lambda^3}+\mathcal{O}\left(\dfrac{\left(m_{3}^{6} + m_{4}^{6} \right) \Lambda^6}{m_1^5 \Lambda^5}\right)\right] \left(\left|S^{34}\right|^2 +\left|S^{43}\right|^2\right) \tag{A.19}, \\
\left[\dfrac{m_{2}^4 \Lambda^4}{960 \pi^2 m_1^3 \Lambda^3}+\mathcal{O}\left(\dfrac{m_{2}^{6} \Lambda^6}{m_1^5 \Lambda^5}\right)\right] \left|S^{22}\right|^2,\tag{A.20} \\
  \left[\dfrac{m_{3}^4 \Lambda^4}{960 \pi^2 m_1^3 \Lambda^3}+\mathcal{O}\left(\dfrac{m_{3}^{6} \Lambda^6}{m_1^5 \Lambda^5}\right)\right] \left|S^{33}\right|^2,\tag{A.21} \\  \left[\dfrac{m_{4}^4 \Lambda^4}{960 \pi^2 m_1^3 \Lambda^3}+\mathcal{O}\left(\dfrac{m_{4}^{6} \Lambda^6}{m_1^5 \Lambda^5}\right)\right] \left|S^{44}\right|^2.\tag{A.22}
\end{gather}
Here, $E^{\prime}$ is defined as
\begin{align}
E^{\prime}= \begin{pmatrix} A & O \\ O & B \end{pmatrix}, \tag{A.23}
\end{align}
with $A, B$ given by
\begin{align}
A= \begin{pmatrix} O & P \\ Q & R
\end{pmatrix},\ B=\begin{pmatrix} O & P^{\ast} \\ Q^{\ast} & R^{\ast} \end{pmatrix}, \tag{A.24} \end{align}
and
\begin{align}
P= \begin{pmatrix} \left(\sqrt{m_1 \Lambda}\right)^{\ast} S_0^T & \bf{0_{3 \times 3}} \\  \bf{0_{3 \times 3}} & \left(\sqrt{m_1 \Lambda}\right)^{\ast} S_0 \end{pmatrix},\ \ Q= \begin{pmatrix} \sqrt{m_1 \Lambda} S_0^{\ast} & \bf{0_{3 \times 3}} \\ \bf{0_{3 \times 3}} & \sqrt{m_1 \Lambda} S_0^{\dagger}\end{pmatrix}, \ \  R=\begin{pmatrix} S_0^{\ast} S_0^T & \bf{0_{3 \times 3}} \\ \bf{0_{3 \times 3}} & S_0^{\dagger} S_0 \end{pmatrix}. \tag{A.25} \end{align}
\\

{\flushleft \ }
\noindent
The dual singlet fermions, i.e. $\psi_{S^{ij}}$ acquire the mass through the effective $K\ddot{a}hler$ potential:
\begin{align}
K_{eff} \supset -\dfrac{1}{2} Tr \int \dfrac{d^4 p}{\left(2 \pi \right)^4} \dfrac{1}{p^2+ \mathcal{M}^{\dagger} \mathcal{M}}=-\dfrac{1}{2} Tr \int \dfrac{d^4 p}{\left(2 \pi \right)^4}\dfrac{1}{p^2+ m_1 \Lambda} \left({\bf{1}_{12 \times 12}} + \dfrac{A}{p^2+ m_1 \Lambda}\right)^{-1}, \tag{A.26}
\end{align}
which leads to the coupling:
\begin{align}
K_{eff} \supset \sum_{i < j} \dfrac{\left|S^{ii}\right|^2+\left|S^{jj} \right|^2}{48 \pi^2 m_1 \Lambda} \left(\left|S^{ij} \right|^2+ \left|S^{ji} \right|^2\right)+ \sum_{i} \dfrac{\left|S^{ii}\right|^4}{96 \pi^2 m_1 \Lambda},  \tag{A.27}
\end{align}
where $i, j$ runs $2-4$.

\newpage

{\begin{center} \bf B.\ Numerical analysis for the generation of the lepton number\end{center}}

\noindent
We here provide the numerical estimation of the lepton asymmetry, taking account of the quasi mass-degeneracy between $\chi^3_{+}$ and $\chi^3_{-}$. Note that the Boltzmann equations take a different form, depending on during or after the reheating stage. For $T>T_{RH}$, the equations involved are expressed as
\begin{gather}
\dfrac{d \left(z^5 N_{\chi^3_+}\right)}{dz} \simeq \dfrac{8}{3} \cdot  D_+ \times \left(z^5 N_{\chi^3_+}^{eq}\right),\tag{B.1} \\
\dfrac{d \left(z^5 N_{\chi^3_-}\right)}{dz} \simeq \dfrac{8}{3} \cdot D_- \times \left(z^5 N_{\chi^3_-}^{eq}\right),\tag{B.2}\\
\dfrac{d \left(z^5N_{B-L}\right)}{dz} \simeq \dfrac{8}{3} \left[\epsilon^{\chi^{3}_{+}}_{l_n} D_+ \times \left(z^5 N_{\chi^3_+}^{eq}\right)+ \epsilon^{\chi^{3}_{-}}_{l_n}D_- \times \left(z^5 N_{\chi^3_-}^{eq}\right)- \left(W_+^{ID}+W_-^{ID}\right) \times \left(z^5 N_{B-L}\right)\right].\tag{B.3}
\end{gather}
On the other hand, for $T<T_{RH}$, we may write the following ones:
\begin{gather}
\dfrac{d N_{\chi^3_+}}{dz} \simeq  D_+ \times N_{\chi^3_+}^{eq},\tag{B.4}\\
\dfrac{d N_{\chi^3_-}}{dz} \simeq  D_- \times N_{\chi^3_-}^{eq},\tag{B.5}\\
\dfrac{d N_{B-L}}{dz} \simeq \epsilon^{\chi^{3}_{+}}_{l_n}D_+ \times N_{\chi^3_+}^{eq}+\epsilon^{\chi^{3}_{-}}_{l_n}D_- \times N_{\chi^3_-}^{eq}- \left(W_+^{ID}+W_-^{ID}\right) \times N_{B-L}. \tag{B.6}
\end{gather}
Here, $D_{\pm}, W_{\pm}^{ID}$ and $z$ are defined as
\begin{equation}
D_{\pm} \equiv K_{\chi^{3}_{\pm}}^{\tilde{h}_u, l_k} \times  \dfrac{z \cdot K_1 \left(z \right)}{K_2 \left(z \right)},\ \ \ W_{\pm}^{ID} \equiv \dfrac{1}{2} \cdot D_{\pm} \times \dfrac{N_{\chi^3_{\pm}}^{eq}}{N^{eq}_l}, \ \ \ z \equiv m_{\chi^3_-}/T \tag{B.7}
\end{equation}
%\begin{equation}
%D_{\pm} \equiv K_{\chi^{3}_{\pm}}^{\tilde{h}_u, l_k} \times z^4, \ \ \text{for}\ \ z<1 \tag{B.8}
%\end{equation}
with $K_1, K_2$ denoting the modified Bessel functions respectively. For simplicity, we neglected the scattering processes concerned. Besides, it should be emphasized that $m_{\chi^3_+}/T \simeq m_{\chi^3_-}/T$ in contrast to the hierarchical right-handed neutrino mass scenario. Noting that Eq.(B.6) is rewritten as follows:
\begin{gather}
\dfrac{d \left(z^5N_{B-L}\right)}{dz} \simeq D_+ \times \left(z^5 N_{\chi^3_+}^{eq}\right) +D_- \times \left(z^5 N_{\chi^3_-}^{eq}\right) - \left(W_+^{ID}+W_-^{ID}\right) \times \left(z^5 N_{B-L}\right)+ 5z^4 \times N_{B-L}, \tag{B.8}
\end{gather}
we derive $N_{B-L}$ as follows:
\begin{align}
z_f^5 \cdot N_{B-L} \left(z=z_f\right) \simeq & \int_{z_I}^{z_{RH}} dz^{\prime}\  \dfrac{8}{3} \left[\epsilon^{\chi^{3}_{+}}_{l_n} D_+\left(z^5 N_{\chi^3_+}^{eq}\right)+ \epsilon^{\chi^{3}_{-}}_{l_n}D_- \times \left(z^5 N_{\chi^3_-}^{eq}\right) \right] \notag \\ & \times \exp{\left[-\int_{z_{RH}}^{z_f} dz^{\prime \prime} \left(W_+^{ID}+W_-^{ID}\right)\right]} \times \left(\dfrac{z_f}{z_{RH}}\right)^5, \tag{B.9}
\end{align}
which is eventually reduced to the expression of the baryon abundance:
\begin{equation}
\dfrac{n_{B}}{s} \left(z=z_f\right) \simeq 8.808 \times 10^{-11} \times \left(\dfrac{\sum_{n=2}^4 \left|y^n_{34} \right|^2}{0.6}\right) \times \left(\dfrac{T_{RH}}{3.0 \times10^{4}\ \text{GeV}}\right)^7 \left(\dfrac{\sin{2\phi_{cp}}}{1.0}\right). \tag{B.10}
\end{equation}
It is understood that the integral in Eq.(B.9) almost converges at $z \sim 10$.
Additionally, we neglected the wash-out terms during the reheating stage, because of $K_{\chi^{3}_{\pm} RH}^{\tilde{h}_u, l_k}$ (that denotes the relevant factor at $T>T_{RH}$) $\simeq \mathcal{O}(10^{-3}) \times K_{\chi^{3}_{\pm}}^{\tilde{h}_u, l_k}$ at $T<T_{RH}$.

\newpage

{\begin{center} \bf C.\ $\psi_{S^{ij}}$ generation and implication on cosmology\end{center}}

\noindent
We address the abundance of the eight fermions and their implications on cosmology, following [18,19].  The Boltzmann equation of relevance is written as
\begin{equation}
\dfrac{d n_{\psi_{S^{ij}}}}{dt} +3Hn_{\psi_{S^{ij}}} =-\left< \sigma_s v \right> \left(n_{\psi_{S^{ij}}} n_{l_n}-n^{eq}_{\psi_{S^{ij}}} n_{l_n}^{eq}\right)- \left< \sigma_t v \right> \left(n_{\psi_{S^{ij}}} n_t-n^{eq}_{\psi_{S^{ij}}} n_t^{eq}\right), \tag{C.1}
\end{equation}
where each of $\left<\sigma_s v \right>$ or $\left<\sigma_t v \right>$ is the total cross section of $l_n \psi_{S^{ij}}  \rightarrow tt, \ t \psi_{S^{ij}}  \rightarrow t l_n  \ (n=1-3)$, while $n_k\ (k=\psi_{S^{ij}}, t, l_n)$ denotes the number density, and $n_k^{eq}$ is that in thermal equilibrium.
Defining $z$ and $N_k, \ N_k^{eq} (k=\psi_{S^{ij}}, t, l_n)$ as
\begin{gather}
z \equiv \dfrac{m_{\psi_{S^{ij}}}}{T}, \tag{C.2} \\
N_k \equiv \dfrac{n_k}{n_{\gamma}},\ \ N_k^{eq} \equiv \dfrac{n_k}{n_{\gamma}}, \tag{C.3}
\end{gather}
one reduces Eq.(C.1) to the form of
\begin{align}
\dfrac{d N_{\psi_{S^{ij}}}}{dz} &= -\left[F_s \cdot \left(\dfrac{N_{\psi_{S^{ij}}}}{N^{eq}_{\psi_{S^{ij}}}}\dfrac{N_l}{N_l^{eq}}-1\right)+F_t \cdot \left(\dfrac{N_{\psi_{S^{ij}}}}{N^{eq}_{\psi_{S^{ij}}}}\dfrac{N_t}{N^{eq}_t}-1\right)\right] N^{eq}_{\psi_{S^{ij}}}, \notag \\ \tag{C.4}
& =-\left[F_s + F_t\right] \left(N_{\psi_{S^{ij}}}-N_{\psi_{S^{ij}}}^{eq}\right).
\end{align}
Here, $n_{\gamma}$ is the number density of photons, whereas $F_{s, t}$ is given by
\begin{gather}
F_s=\dfrac{\left< \sigma_s v \right> n^{eq}_{\psi_{S^{ij}}}}{H z}=\dfrac{\gamma^{eq}_{s}}{n^{eq}_{\psi_{S^{ij}}} \cdot H z}, \tag{C.5}\\
 F_t=\dfrac{\left< \sigma_t v \right> n^{eq}_{\psi_{S^{ij}}}}{H z}=\dfrac{\gamma^{eq}_{t}}{n^{eq}_{\psi_{S^{ij}}} \cdot H z}. \tag{C.6}
\end{gather}
$\gamma^{eq}_{s, t}$, which denote the reaction density for the processes $l_n \psi_{S^{ij}} \rightarrow tt, \ t \psi_{S^{ij}} \rightarrow tl_n$ as well, are written as follows:
\begin{gather}
\gamma^{eq}_{s}=\dfrac{T^4}{64 \pi^4} \int_{\frac{4 m^2_t}{T^2}}^{\infty} \hat{\sigma}_{s} \sqrt{x} \cdot K_1 \left(\sqrt{x}\right) dx \times \dfrac{1}{z^2 K_2\left(z \right)}, \tag{C.7} \\
\gamma^{eq}_{t}= \dfrac{T^4}{64 \pi^4} \int_{\frac{(m_t+m_{\psi_{S^{ij}}})^2}{T^2}}^{\infty} \hat{\sigma}_{t} \sqrt{x} \cdot K_1 \left(\sqrt{x}\right) dx \times \dfrac{1}{z^2 K_2\left(z \right)} \notag \\ \ \ \ \ \ \simeq \dfrac{T^4}{64 \pi^4} \int_{\frac{m^2_t}{T^2}}^{\infty} \hat{\sigma}_{t} \sqrt{x} \cdot K_1 \left(\sqrt{x}\right) dx \times \dfrac{1}{z^2 K_2\left(z \right)}, \tag{C.8}
\end{gather}
where $K_1, K_2$ are the modified Bessel functions respectively, and we have taken account of $m_{\psi_{S^{ij}}}\ll m_t$. $\hat{\sigma}_{s, t}$ are then given by
\begin{gather}
\hat{\sigma}_s= \dfrac{8}{s} \cdot \left[\left(p_{\psi_{S^{ij}}} \cdot p_{l_n}\right)^2-m^2_{\psi_{S^{ij}}} m^2_{l_n}\right] \cdot \sigma_s \simeq \dfrac{x^{1/2} \left(x-\dfrac{4 m^2_t}{T^2}\right)^{3/2}}{\left(x-\dfrac{m^2_{H^0}}{T^2}\right)^2}, \tag{C.9}
\end{gather}
\begin{align}
& \hat{\sigma}_t = \dfrac{8}{s} \cdot \left[\left(p_{\psi_{S^{ij}}} \cdot p_{t}\right)^2-m^2_{\psi_{S^{ij}}} m^2_{t}\right] \cdot \sigma_t \simeq \dfrac{1}{x^2}  \left[\dfrac{2\left(x-\dfrac{m^2_t}{T^2}\right)^2\left(\left(x-\dfrac{m^2_t}{T^2}\right)^2 +\dfrac{2m^2_{H^0}x}{T^2}\right)}{2\left(x-\dfrac{m^2_t}{T^2}\right)^2+\dfrac{2m^2_{H^0}x}{T^2}} \right] \notag \\
& \ \ \ \ \ \ +\dfrac{1}{x^2} \cdot \dfrac{2m^2_{H^0}x}{T^2} \left[\log{\dfrac{2m^2_{H^0}x}{T^2}}-\log\left({2\left(x-\dfrac{m^2_t}{T^2}\right)^2+\dfrac{2m^2_{H^0}x}{T^2}}\right) \right], \tag{C.10}
\end{align}
where $x \equiv s/T^2$ and $s$ is the squared center of mass energy, while $\sigma_{s, t}$ are the usual cross sections for each process.
Besides, we set $g_\ast=106.75$, only taking into consideration the SM particles. 
We thus integrate Eq.(C.4) using a new variable $z_t \equiv (m_t/m_{\psi_{S^{ij}}}) \cdot z$ (instead of $z$) to obtain $N_{\psi_{S^{ij}}}$:
\begin{gather}
N_{\psi_{S^{ij}}} \simeq   N^{eq}_{\psi_{S^{ij}}} \left[1- \exp{\left(-\mathcal{I}\right)}\right] \simeq \dfrac{3}{4} \left[1- \exp{\left(-\mathcal{I}\right)}\right], \ \ \text{for}\ z \ll 1, \tag{C.11}
\end{gather}
where $\mathcal{I}$ is expressed as
\begin{align}
\mathcal{I} &\equiv \int_0^{z_t} \dfrac{K_S}{6} \cdot \dfrac{m_{\psi_{S^{ij}}}}{m_t} \left(\int_{4 z_t^{\prime 2}}^{\infty} \dfrac{\hat{\sigma}_s}{z^2 K_2\left(z \right)} dx+ 2 \int_{z_t^{\prime 2}}^{\infty} \dfrac{\hat{\sigma}_t}{z^2 K_2\left(z \right)} dx\right) dz^{\prime}_t \notag \\ 
&\ \simeq \int_0^{z_t} \dfrac{K_S}{12}\cdot \dfrac{m_{\psi_{S^{ij}}}}{m_t} \left(\int_{4 z_t^{\prime 2}}^{\infty} \hat{\sigma}_s dx+ 2 \int_{z_t^{\prime 2}}^{\infty} {\sigma}_t dx\right) dz^{\prime}_t \tag{C.12}
\end{align}
with
\begin{gather}
K_S \equiv \sum_{n=1}^3 \left(y_{ij}^n \sin{\beta}\right)^2 \left(\dfrac{\Lambda}{M_{pl}}\right)^2 \dfrac{v^2}{m_{\psi_{S^{ij}}}} \cdot \dfrac{1}{m_{\ast}^s}, \tag{C.13} \\
m_{\ast}^s \equiv \dfrac{4 \pi^2}{9} \cdot \dfrac{2 v^2}{m_t} \times \dfrac{16 \pi^{5/2} \sqrt{g_{\ast}}}{3 \sqrt{5}} \cdot \dfrac{v^2}{\sqrt{8 \pi} M_{pl}}. \tag{C.14} 
\end{gather}
Here, it is understood that $z^2 K_2\left(z \right)\simeq 2$ for $z \ll 1$. Regarding the derivation of $K_S$, we allowed for the interactions among the Lagrangian density:
\begin{gather}
\mathcal{L} \supset \sum_{n=1}^3 \ \sum_{i, j=2}^4 y_{ij}^n\dfrac{\Lambda S^{ij} H_u L_n}{M_{pl}}+Y_t H_u t_L \bar{t}_R \supset \sum_{n=1}^3 \ \sum_{i, j=2}^4 y_{ij}^n \sin{\beta} \cdot \dfrac{\Lambda S^{ij} H^0 L_n}{M_{pl}}+ Y_t \sin{\beta} H^0 t_L \bar{t}_R. \tag{C.15}
\end{gather}Taking into consideration that $\mathcal{I}$ almost converges at $z_{t} \sim 3.5$, i.e. $T \sim 50$ GeV, each of $N_{\psi_{S^{ij}}}$ is roughly estimated as
\begin{gather}
N_{\psi_{S^{34}}} \simeq N_{\psi_{S^{43}}} \simeq 1.11\times10^{-1}, \tag{C.16}\\ %0.11 y multiplied by sqrt2
N_{\psi_{S^{ij}}}^{OT} \simeq 1.50 \times 10^{-2}. \tag{C.17}
\end{gather}
Here, the superscript $OT$  of Eq.(C.17) denotes the other fermions, i.e. $(i, j)=(2, 3), (2, 4),$ $(3, 2), (4, 2)$. Subsequently, we evaluate their energy density at the decay temperature, compared with the radiation energy density. Noting that the temperature of $\psi_{S^{ij}}$, after decoupling the thermal bath, is given by
\begin{equation}
T^{\prime}_{\psi_{S^{ij}}} = \left(\dfrac{10.75}{g_{\ast s}}\right)^{1/3} T = 0.475 \times \left(\dfrac{100}{g_{\ast s}}\right)^{1/3}T,\tag{C.18}
\end{equation}
we obtain the relevant ratios:
\begin{align}
\left[\dfrac{\rho_{{\psi_{S^{34}}}}+ \rho_{{\psi_{S^{43}}}}}{\rho_{rad.}}\right]_{T={T^{decay}_{\psi_{S^{34}}}}}  &\simeq \left[\left(N_{\psi_{S^{34}}} m_{\psi_{S^{34}}} + N_{\psi_{S^{43}}} m_{\psi_{S^{43}}}\right) \cdot \dfrac{\left(T_{\psi_{S^{ij}}}^{\prime}\right)^{3}}{\rho_{rad.}} \cdot \dfrac{g_{\psi_{S^{ij}}} \zeta \left(3 \right)}{\pi^2}\right]_{T={T^{decay}_{\psi_{S^{34}}}}}  \notag \\ &\simeq \dfrac{60 \cdot \zeta \left(3 \right)}{\pi^4} \cdot \dfrac{1}{g_{\ast s}} \cdot \dfrac{N_{\psi_{S^{34}}} m_{\psi_{S^{34}}} + N_{\psi_{S^{43}}} m_{\psi_{S^{43}}}}{T^{decay}_{\psi_{S^{34}}}} \simeq 0.122,  \tag{C.19}\\  %0.118
\left[\dfrac{\sum_{OT} \rho_{{\psi_{S^{ij}}}}}{\rho_{rad.}}\right]_{T=T^{OT\ decay}_{\psi_{S^{ij}}}} &\simeq  \left[\left(\sum_{OT}N_{\psi_{S^{ij}}} m_{\psi_{S^{ij}}}\right) \cdot \dfrac{\left(T_{\psi_{S^{ij}}}^{\prime}\right)^3}{\rho_{rad.}} \cdot \dfrac{g_{\psi_{S^{ij}}} \zeta \left(3 \right)}{\pi^2}\right]_{T=T^{OT\ decay}_{\psi_{S^{ij}}}} \notag \\ 
& \simeq \dfrac{60 \cdot \zeta \left(3 \right)}{\pi^4} \cdot \dfrac{1}{g_{\ast s}} \cdot \left(\sum_{OT}N_{\psi_{S^{ij}}} m_{\psi_{S^{ij}}}\right) \cdot \dfrac{1}{T^{OT\ decay}_{\psi_{S^{ij}}}}\simeq 9.25 \times 10^{-2}, \tag{C.20} %0.122
\end{align}
where we set $g_{\ast s}=100$ because the decoupling temperature is around $50$ GeV.

\begin{thebibliography}{999}
\bibitem{planck-suppressed-dual-mass}
The following couplings are present, though those are irrelevant in our model.
\begin{equation}
\sum_{i,j=2}^4 \kappa_{ij} \dfrac{\bar{Q}^i Q^1 \bar{Q}^jQ^1}{M_{pl}}+ \eta_{ij} \dfrac{\bar{Q}^1 Q^i \bar{Q}^1 Q^j}{M_{pl}}.
\end{equation}
\bibitem{S-two-phi} 
In the presence of the $\kappa  \bar{Q}^iQ^j \bar{Q}^kQ^l/M_{pl}$, the coefficient $\kappa$ would need to be highly fine-tuned:
\begin{equation}
\left|\kappa \right| \lesssim 10^{-7} \left|y^n_{ij} \right| \notag
\end{equation}
so that the dual singlet bosons are not thermalized due to their prompt decay.
\bibitem{note1}
For complimentary argument, we stress that even though the MSSM squarks are likely to be non-relativistic as the Higgs mass of 125 GeV requires the SM stop to be heavier than $10^4$ GeV, the dual singlet bosons can be generated from the inverse decay of the (other) MSSM particles.
\end{thebibliography}
\end{document}